
\font\elevenrm=cmr10  at 11pt
\font\tenrm=cmr10
\font\ninerm=cmr9
\font\eightrm=cmr8
\font\sevenrm=cmr7
\font\sixrm=cmr6


\font\twelvemi=cmmi12
\font\elevenmi=cmmi10  at 11pt

\font\eightmi=cmmi8

\font\sixmi=cmmi6
\font\fivemi=cmmi5
\skewchar\twelvemi='177 \skewchar\eightmi='177 \skewchar\sixmi='177
    \skewchar\fivemi='177
\skewchar\elevenmi='177 \skewchar\eightmi='177 \skewchar\sixmi='177
    \skewchar\fivemi='177


\font\twelvesy=cmsy10 at 12pt
\font\elevensy=cmsy10 at 11pt

\font\ninesy=cmsy9
\font\eightsy=cmsy8
\font\sevensy=cmsy7
\font\sixsy=cmsy6

\skewchar\twelvesy='60 \skewchar\eightsy='60 \skewchar\sixsy='60
\skewchar\elevensy='60 \skewchar\eightsy='60 \skewchar\sixsy='60


\font\elevenbf=cmbx10 at 11pt

\font\eightbf=cmbx8
\font\sixbf=cmbx6

\newfam\eufam


\font\twelvett=cmtt12
\hyphenchar\twelvett=-1 
\font\eleventt=cmtt10 at 11pt

\hyphenchar\eleventt=-1 


\font\elevensl=cmsl10 at 11pt

\newfam\cmcscfam

\font\elevencmcsc=cmcsc10 at 11pt


\font\elevenit=cmti10 at 11pt

\newfam\msbmfam

\font\elevenmsbm=msbm10 at 11pt

\font\eightmsbm=msbm8

\font\sixmsbm=msbm6

\newfam\msamfam

\newfam\cmssfam

\font\elevencmss=cmss10 at 11pt

\font\ninecmss=cmss9
\font\eightcmss=cmss8


\font\elevenex=cmex10 at 11pt

\catcode`@=11
\def\elevenpoint{\def\rm{\fam0\elevenrm}%
   \textfont0=\elevenrm \scriptfont0=\eightrm \scriptscriptfont0=\sixrm
   \textfont1=\elevenmi \scriptfont1=\eightmi \scriptscriptfont1=\sixmi
   \textfont2=\elevensy \scriptfont2=\eightsy \scriptscriptfont2=\sixsy
   \textfont3=\elevenex \scriptfont3=\elevenex
\scriptscriptfont3=\elevenex
  \def\it{\fam\itfam\elevenit}%
    \textfont\itfam=\elevenit
  \def\sl{\fam\slfam\elevensl}%
    \textfont\slfam=\elevensl
  \def\bf{\fam\bffam\elevenbf}%
    \textfont\bffam=\elevenbf
    \scriptfont\bffam=\eightbf
    \scriptscriptfont\bffam=\sixbf
  \def\cmcsc{\fam\cmcscfam\elevencmcsc}%
    \textfont\cmcscfam=\elevencmcsc
   \def\msbm{\fam\msbmfam\elevenmsbm}%
     \textfont\msbmfam=\elevenmsbm \scriptfont\msbmfam=\eightmsbm
     \scriptscriptfont\msbmfam=\sixmsbm
  \def\cmss{\fam\cmssfam\elevencmss}%
    \textfont\cmssfam=\elevencmss \scriptfont\cmssfam=\ninecmss
    \scriptscriptfont\cmssfam=\eightcmss
  \normalbaselineskip=13.2pt
  \smallskipamount=3.3pt plus 1.1pt minus 1.1pt
  \medskipamount=  6.6pt plus 2.2pt minus 2.4pt
  \bigskipamount= 13.2pt plus 4.4pt minus 4.4pt
  \parskip=1pt plus .5pt
  \normallineskip=1.5pt
  \normallineskiplimit=1.5pt
  \widowpenalty=10000
  \clubpenalty=10000
  \setbox\strutbox=\hbox{\vrule height9.35pt depth3.85pt width0pt}%
  \def\big##1{{\hbox{$\left##1\vbox to9.35\p@{}\right.\n@space$}}}%
  \def\Big##1{{\hbox{$\left##1\vbox to12.65\p@{}\right.\n@space$}}}%
  \def\bigg##1{{\hbox{$\left##1\vbox to15.95\p@{}\right.\n@space$}}}%
  \def\Bigg##1{{\hbox{$\left##1\vbox to19.25\p@{}\right.\n@space$}}}%
  \normalbaselines\rm}%
\catcode`@=12

\expandafter\ifx\csname pre amssym.tex at\endcsname\relax
\else\endinput\fi
\expandafter\chardef\csname pre amssym.tex at\endcsname=\the\catcode`\@
\catcode`\@=11
\ifx\undefined\newsymbol \else \begingroup\def\input#1 {\endgroup}\fi
\input amssym.def \relax
\newsymbol\boxdot 1200
\newsymbol\boxplus 1201
\newsymbol\boxtimes 1202
\newsymbol\square 1003
\newsymbol\blacksquare 1004
\newsymbol\centerdot 1205
\newsymbol\lozenge 1006
\newsymbol\blacklozenge 1007
\newsymbol\circlearrowright 1308
\newsymbol\circlearrowleft 1309
\undefine\rightleftharpoons
\newsymbol\rightleftharpoons 130A
\newsymbol\leftrightharpoons 130B
\newsymbol\boxminus 120C
\newsymbol\Vdash 130D
\newsymbol\Vvdash 130E
\newsymbol\vDash 130F
\newsymbol\twoheadrightarrow 1310
\newsymbol\twoheadleftarrow 1311
\newsymbol\leftleftarrows 1312
\newsymbol\rightrightarrows 1313
\newsymbol\upuparrows 1314
\newsymbol\downdownarrows 1315
\newsymbol\upharpoonright 1316
 
\newsymbol\downharpoonright 1317
\newsymbol\upharpoonleft 1318
\newsymbol\downharpoonleft 1319
\newsymbol\rightarrowtail 131A
\newsymbol\leftarrowtail 131B
\newsymbol\leftrightarrows 131C
\newsymbol\rightleftarrows 131D
\newsymbol\Lsh 131E
\newsymbol\Rsh 131F
\newsymbol\rightsquigarrow 1320
\newsymbol\leftrightsquigarrow 1321
\newsymbol\looparrowleft 1322
\newsymbol\looparrowright 1323
\newsymbol\circeq 1324
\newsymbol\succsim 1325
\newsymbol\gtrsim 1326
\newsymbol\gtrapprox 1327
\newsymbol\multimap 1328
\newsymbol\therefore 1329
\newsymbol\because 132A
\newsymbol\doteqdot 132B
 
\newsymbol\triangleq 132C
\newsymbol\precsim 132D
\newsymbol\lesssim 132E
\newsymbol\lessapprox 132F
\newsymbol\eqslantless 1330
\newsymbol\eqslantgtr 1331
\newsymbol\curlyeqprec 1332
\newsymbol\curlyeqsucc 1333
\newsymbol\preccurlyeq 1334
\newsymbol\leqq 1335
\newsymbol\leqslant 1336
\newsymbol\lessgtr 1337
\newsymbol\backprime 1038
\newsymbol\risingdotseq 133A
\newsymbol\fallingdotseq 133B
\newsymbol\succcurlyeq 133C
\newsymbol\geqq 133D
\newsymbol\geqslant 133E
\newsymbol\gtrless 133F
\newsymbol\sqsubset 1340
\newsymbol\sqsupset 1341
\newsymbol\vartriangleright 1342
\newsymbol\vartriangleleft 1343
\newsymbol\trianglerighteq 1344
\newsymbol\trianglelefteq 1345
\newsymbol\bigstar 1046
\newsymbol\between 1347
\newsymbol\blacktriangledown 1048
\newsymbol\blacktriangleright 1349
\newsymbol\blacktriangleleft 134A
\newsymbol\vartriangle 134D
\newsymbol\blacktriangle 104E
\newsymbol\triangledown 104F
\newsymbol\eqcirc 1350
\newsymbol\lesseqgtr 1351
\newsymbol\gtreqless 1352
\newsymbol\lesseqqgtr 1353
\newsymbol\gtreqqless 1354
\newsymbol\Rrightarrow 1356
\newsymbol\Lleftarrow 1357
\newsymbol\veebar 1259
\newsymbol\barwedge 125A
\newsymbol\doublebarwedge 125B
\undefine\angle
\newsymbol\angle 105C
\newsymbol\measuredangle 105D
\newsymbol\sphericalangle 105E
\newsymbol\varpropto 135F
\newsymbol\smallsmile 1360
\newsymbol\smallfrown 1361
\newsymbol\Subset 1362
\newsymbol\Supset 1363
\newsymbol\Cup 1264
 
\newsymbol\Cap 1265
 
\newsymbol\curlywedge 1266
\newsymbol\curlyvee 1267
\newsymbol\leftthreetimes 1268
\newsymbol\rightthreetimes 1269
\newsymbol\subseteqq 136A
\newsymbol\supseteqq 136B
\newsymbol\bumpeq 136C
\newsymbol\Bumpeq 136D
\newsymbol\lll 136E
 
\newsymbol\ggg 136F
 
\newsymbol\circledS 1073
\newsymbol\pitchfork 1374
\newsymbol\dotplus 1275
\newsymbol\backsim 1376
\newsymbol\backsimeq 1377
\newsymbol\complement 107B
\newsymbol\intercal 127C
\newsymbol\circledcirc 127D
\newsymbol\circledast 127E
\newsymbol\circleddash 127F
\newsymbol\lvertneqq 2300
\newsymbol\gvertneqq 2301
\newsymbol\nleq 2302
\newsymbol\ngeq 2303
\newsymbol\nless 2304
\newsymbol\ngtr 2305
\newsymbol\nprec 2306
\newsymbol\nsucc 2307
\newsymbol\lneqq 2308
\newsymbol\gneqq 2309
\newsymbol\nleqslant 230A
\newsymbol\ngeqslant 230B
\newsymbol\lneq 230C
\newsymbol\gneq 230D
\newsymbol\npreceq 230E
\newsymbol\nsucceq 230F
\newsymbol\precnsim 2310
\newsymbol\succnsim 2311
\newsymbol\lnsim 2312
\newsymbol\gnsim 2313
\newsymbol\nleqq 2314
\newsymbol\ngeqq 2315
\newsymbol\precneqq 2316
\newsymbol\succneqq 2317
\newsymbol\precnapprox 2318
\newsymbol\succnapprox 2319
\newsymbol\lnapprox 231A
\newsymbol\gnapprox 231B
\newsymbol\nsim 231C
\newsymbol\ncong 231D
\newsymbol\diagup 201E
\newsymbol\diagdown 201F
\newsymbol\varsubsetneq 2320
\newsymbol\varsupsetneq 2321
\newsymbol\nsubseteqq 2322
\newsymbol\nsupseteqq 2323
\newsymbol\subsetneqq 2324
\newsymbol\supsetneqq 2325
\newsymbol\varsubsetneqq 2326
\newsymbol\varsupsetneqq 2327
\newsymbol\subsetneq 2328
\newsymbol\supsetneq 2329
\newsymbol\nsubseteq 232A
\newsymbol\nsupseteq 232B
\newsymbol\nparallel 232C
\newsymbol\nmid 232D
\newsymbol\nshortmid 232E
\newsymbol\nshortparallel 232F
\newsymbol\nvdash 2330
\newsymbol\nVdash 2331
\newsymbol\nvDash 2332
\newsymbol\nVDash 2333
\newsymbol\ntrianglerighteq 2334
\newsymbol\ntrianglelefteq 2335
\newsymbol\ntriangleleft 2336
\newsymbol\ntriangleright 2337
\newsymbol\nleftarrow 2338
\newsymbol\nrightarrow 2339
\newsymbol\nLeftarrow 233A
\newsymbol\nRightarrow 233B
\newsymbol\nLeftrightarrow 233C
\newsymbol\nleftrightarrow 233D
\newsymbol\divideontimes 223E
\newsymbol\varnothing 203F
\newsymbol\nexists 2040
\newsymbol\Finv 2060
\newsymbol\Game 2061
\newsymbol\mho 2066
\newsymbol\eth 2067
\newsymbol\eqsim 2368
\newsymbol\beth 2069
\newsymbol\gimel 206A
\newsymbol\daleth 206B
\newsymbol\lessdot 236C
\newsymbol\gtrdot 236D
\newsymbol\ltimes 226E
\newsymbol\rtimes 226F
\newsymbol\shortmid 2370
\newsymbol\shortparallel 2371
\newsymbol\smallsetminus 2272
\newsymbol\thicksim 2373
\newsymbol\thickapprox 2374
\newsymbol\approxeq 2375
\newsymbol\succapprox 2376
\newsymbol\precapprox 2377
\newsymbol\curvearrowleft 2378
\newsymbol\curvearrowright 2379
\newsymbol\digamma 207A
\newsymbol\varkappa 207B
\newsymbol\Bbbk 207C
\newsymbol\hslash 207D
\undefine\hbar
\newsymbol\hbar 207E
\newsymbol\backepsilon 237F
\catcode`\@=\csname pre amssym.tex at\endcsname



\input epsf

\newcount\equationno
\def\Eqn#1/{%
  \csname EQ#1\endcsname}
\def\setEqn#1/#2/{%
  \expandafter\xdef\csname EQ#1\endcsname{#2}}
\def\N/{%
  \global\advance\equationno 1
  \the\equationno}
\def\SN#1/{%
  \global\advance\equationno 1
  \expandafter\xdef\csname EQ#1\endcsname{\the\equationno}%
  \Eqn#1/}
\def\SNM#1/{%
  \SN#1/%
  \ifproofmode
    \rlap{\hskip .25in\sevenrm #1}%
  \fi}
\def\NextN{{\advance \equationno 1 \the\equationno}}
\def\NextNN{{\advance \equationno 2 \the\equationno}}
\def\NextNNN{{\advance \equationno 3 \the\equationno}}

\newcount\smallpenalty \smallpenalty=-50
\newcount\medpenalty   \medpenalty=-100
\newcount\bigpenalty   \bigpenalty=-200

\catcode`@=11
\def\ninebig#1{{\hbox{$\textfont0=\ninerm \textfont1=\ninei
\textfont2=\ninesy
  \left#1\vbox to7.25pt{}\right.\n@space$}}}
\catcode`@=12

\catcode`@=11
\def\sevenbig#1{{\hbox{$\textfont0=\sevenrm
  \textfont1=seveni \textfont2=\sevensy
  \left#1\vbox to 6pt{}\right.\n@space$}}}
\catcode`@=12


\def\TwoDigits#1{%
   \ifnum \the #1 < 10
      0%
   \fi
   \the #1}


\catcode`@=11
\def\leftmathdisplay#1{\displ@y\halign {%
   \hskip\parindent $\displaystyle{##}$\hfil&&
   \quad $\displaystyle{{}##}$\hfil\crcr #1\crcr}}

\newdimen \displayshortwidth
\def\setdisplayshortwidth{%
        \displayshortwidth=\displaywidth
        \advance\displayshortwidth -\parindent}

\def\blackbox{\vrule height 6pt width4pt depth0pt}

\outer\def\proclaim #1. #2\par{\medbreak
  \noindent{\bold #1.\enspace}{\sl#2\par}%
  \ifdim\lastskip<\medskipamount \removelastskip\penalty55\medskip\fi}

\def\Not#1{%
  \setbox0=\hbox{$#1$}%
  \mathrel{\rlap{$#1$}\hbox to
\wd0{\hfil$\not\mathrel{\hphantom{=}}$\hfil}}}

\newif\ifproofmode
\newif\ifexpressmode

\newcount\equationno
\def\Eqn#1/{%
  \csname EQ#1\endcsname}
\def\setEqn#1/#2/{%
  \expandafter\xdef\csname EQ#1\endcsname{#2}}
\def\N/{%
  \global\advance\equationno 1
  \the\equationno}
\def\SN#1/{%
  \global\advance\equationno 1
  \expandafter\xdef\csname EQ#1\endcsname{\the\equationno}%
  \Eqn#1/}
\def\SNM#1/{%
  \SN#1/%
  \ifproofmode
    \rlap{\hskip .25in\sevenrm #1}%
  \fi}
\def\NextN{{\advance \equationno 1 \the\equationno}}
\def\NextNN{{\advance \equationno 2 \the\equationno}}
\def\NextNNN{{\advance \equationno 3 \the\equationno}}

\newcount \theoremno
\def\nexttheorem{%
  \global\advance\theoremno 1
  \the\theoremno}
\newdimen \oldparindent
\oldparindent = \parindent

\equationno=0
\def\sbullet{{\scriptscriptstyle{+}}}
\def\bu{\sbullet}
\def\IV{{\ninerm IV} }
\catcode`@=11
\def\fat{%
        \relax
        \ifmmode
                \expandafter\mathpalette
                \expandafter\f@tmath
        \else
                \expandafter\f@tord
        \fi}
\def\f@tord#1{%
        \setbox\z@=\hbox{#1}%
        \finishf@t}
\def\f@tmath#1#2{%
        \setbox\z@=\hbox{$\m@th#1{#2}$}%
        \finishf@t}
\def\finishf@t{%
        \kern-.025em\copy\z@\kern-\wd\z@
        \kern.040em\copy\z@\kern-\wd\z@
        \kern-.015em\box\z@}
\catcode`@=12
\def\beginsection#1#2{\vskip0pt plus.1\vsize\penalty-250
    \vskip0pt plus-.1\vsize\smallskip\smallskip\vskip\parskip
     \message{#1#2}\noindent{{\rm #1} {\it#2}}\par\nobreak\smallskip\indent}
\def\beginref{\par\bgroup
    \parindent=0pt\everypar={\hangindent=36pt\hangafter=1}}
\def\endref{\par\egroup}

\def\R{{\Bbb R}}

\def\Larrow{\buildrel {\cal L}\over \rightarrow}

\def\cov{\mathop{\rm cov}\nolimits}

\def\hat{\widehat}
\def\ip<#1>{\langle #1 \rangle} 
\def\notSigma{\Sigma\kern-8pt \bigm|}
\def\varphi{\phi}
\def\tilde{\widetilde}
\def\frac#1#2{{\textstyle{#1\over #2}}}

\def\var{\mathop{\rm var}\nolimits}

\def\Kvee{{\breve K}}
\def\Vvee{{\breve V}}
\def\Larrow{\buildrel {\cal L}\over \rightarrow}
\def\blackbox{\vrule height8pt width0.4pt depth0pt
   \kern-0.4pt\hbox to 8pt{\hrulefill}
     \kern-3.5pt\vrule height8pt width0.4pt depth0pt
     \kern-8.2pt\raise8pt\hbox to 8.3pt{\hrulefill}}
\def\Corollary#1. #2\par{%
   \medbreak{\elevencmcsc Corollary #1.\enspace}{\sl#2}}

\def\Example.{{\elevencmcsc Example.\enspace}}
\def\Lemma#1. #2\par{%
    \medbreak{\elevencmcsc Lemma #1.\enspace}{\sl#2}}

\def\Proof.{{\elevencmcsc Proof.\enspace}}

\def\ProofOfLemma#1.{{\elevencmcsc Proof of Lemma #1.\enspace}}
\def\ProofOfTheorem#1.{{\elevencmcsc Proof of Theorem #1.\enspace}}

\def\ProofOfProposition#1.{{\elevencmcsc Proof of Proposition #1.\enspace}}
\def\Proposition#1. #2\par{%
    \medbreak{\elevencmcsc Proposition #1.\enspace}{\sl#2}}

\def\Theorem#1. #2\par{%
   \medbreak{\elevencmcsc Theorem #1.\enspace}{\sl#2}}
\def\TheoremEnd{\null\nobreak\hfill\blackbox\par\medbreak}
\def\footer{\footline={\hss\tenrm\folio\hss}}
\hsize=6.5truein
\voffset=24truept
\advance\vsize by -\voffset
\parindent=28pt
\overfullrule=0pt
\elevenpoint\rm
\nopagenumbers

\def\foot{
     }
\def\other#1{\ninerm #1}
\def\nsf#1{\ninerm This work was supported in part by #1.}
\def\grantfootnote#1{\parindent=28pt
		     \everypar={\hangindent=28pt\hangafter=1}
   	             \footnote*{\ninerm%
        \edef\grantno{#1}\ifx\grantno\empty
     \else
        \other{#1}
     \fi
        \foot
     }}

\def\statistics{%
   $$\vbox{%
     \halign{\hfil##\hfil\cr
     $^1$Department of Statistics\cr
     The University of Chicago\cr
     Chicago, Illinois  60637\cr
\noalign{\vskip5pt}
     $^2$Department of Physics\cr
	Carthage College\cr
	Kenosha, WI 53140\cr
     }}
   $$
  }

\def\author{\centerline{\sl\name}}
\def\title{\bf Estimating the $K$ function of a point process with}
\def\titleB{\bf an application to cosmology}
\def\name{Michael L.~Stein$^1$, Jean M.~Quashnock$^{1,2}$ and Ji Meng Loh$^1$}
\def\technicalreportno{\vfill\centerline{TECHNICAL REPORT NO.
\trnumber}\vfill}
\def\date{\centerline{\day}}
\def\revision{\smallskip\centerline{\sl Revised\/\ \revisiondate}}

\def\day{March 1999}
\def\revisiondate{May 2000}
\def\trnumber{485}

\normalbaselines
\centerline{\title}
\centerline{\titleB%
\grantfootnote{\nsf{National Science Foundation
                    grants DMS 95-04470 and 99-71127 (Loh and Stein)  
		and NASA grant NAG 5-4406 and NSF grant DMS 97-09696
		(Quashnock)
                }\hfil\smallbreak}}

\bigskip
\author
\technicalreportno
\statistics
\date
\revision

\vfill\eject

\centerline{\bf\title}
\centerline{\bf\titleB%
\grantfootnote{\nsf{National Science Foundation
                    grant DMS 95-04470 and 99-71127 (Loh and Stein)
                and NASA grant NAG 5-4406 and NSF grant DMS 97-09696
                (Quashnock) }\hfil\smallbreak}}
\bigskip
\centerline{Michael L.~Stein, Jean M.~Quashnock and Ji Meng Loh}
\vfill
{\baselineskip=24pt plus .5pt
\narrower\noindent
{\bf Abstract:}
Motivated by the study of an important data set for understanding
the large-scale structure of the universe, this work considers
the estimation of the reduced second moment function, or $K$ function,
of a stationary point process on $\R$ observed over a large number
of segments of possibly varying lengths.
Theory and simulation are used to compare the behavior of isotropic
and rigid motion correction estimators and some modifications of
these estimators.
These results generally support the use of modified versions of
the rigid motion correction.
When applied to a catalog of astronomical objects known as absorbers,
the proposed methods confirm results from earlier analyses
of the absorber catalog
showing clear evidence of clustering up to 50 $h^{-1}$ Mpc 
and marginal evidence for clustering of matter
on spatial scales beyond 100 $h^{-1}$ Mpc, which is beyond the
distance at which clustering of matter is now generally accepted
to exist.
\vfill
\par\noindent
Key words: Reduced second moment function; bootstrapping; large-scale
	structure of the universe; heavy-element absorption-line systems 
\par\noindent
AMS 1991 subject classifications:  Primary 62M30; secondary 62P35, 60G55.
\par\noindent 
Running title: Estimation for point processes

\par}
\vfill
\eject

\clubpenalty=10000
\widowpenalty=10000
\parskip=3pt plus 1pt
\baselineskip=24pt plus .5pt
\lineskip=4pt plus 2pt
\lineskiplimit=4pt
\abovedisplayskip=13pt plus 3pt minus 2pt
\belowdisplayskip=13pt plus 3pt minus 2pt
\pageno=1
\footer

\beginsection{1.}{Introduction}
One way to describe a stationary 
spatial point processes is through some measure
of clumpiness of the events of the process.
A commonly used measure of clumpiness is the reduced
second moment function $K(t)$, defined
as the expected number of events within distance $t$ of a typical
event of the process divided by the intensity of the process.
For a homogeneous Poisson process on $\R^d\! ,$ $K(t)=\mu_d t^d\! ,$
where $\mu_d$ is the volume of a unit ball in $d$ dimensions.
Thus, values of $K(t)$ greater than $\mu_d t^d$ are indicative
of a process that is clumpier than Poisson and values less than
$\mu_d t^d$ are indicative of a process that is more regular than
Poisson.
When estimating $K(t)$
based on observing a process within a bounded window $W$, a central
problem is that for any event in $W$ that is within $t$ of
the boundary of $W$, we do not know for sure how many other
events are within $t$ of it.
Baddeley (1998) describes a number of ways of accounting for these
edge effects.
Although there is
quite a bit of asymptotic theory for how these estimators
behave when the underlying process is Poisson 
(Ripley 1988, Stein 1993),
much less is known for non-Poisson processes.

An interesting aspect of asymptotic theory for point processes is
how one should take limits.
Ripley (1988) and Stein (1993) consider a single growing window,
which might appear to be the obvious way to take limits.
However,
Baddeley, et al.\ (1993) describe applications in which point processes
are observed in many well-separated windows.
For this setting, Baddeley and Gill (1997) argue that it is natural
to consider taking limits by keeping the size of these windows
fixed and letting their number increase.
As they point out, one advantage of this approach is that the edge
effects do not become negligible in the limit, since for any fixed
$t$, the fraction of events that are within $t$ of a window boundary
does not tend to 0.
Thus, for comparing different approaches for handling edge effects,
increasing the number of windows may be more informative than
allowing a single region to grow in all dimensions, for which
the fraction of events that are within $t$ of a window boundary
does tend to 0.
Another advantage of taking limits by letting the number
of windows increase is that if the process is independent
in different regions, then limit theorems are easier to prove.
This is particularly the  case when the windows are all 
well-separated translations of
the same set so that the observations of the process on the multiple windows
can be reasonably modeled as iid realizations.
Baddeley and Gill (1997) use this approach to obtain weak convergence
results for estimators of $K$ and other functions describing
point process behavior.
The resulting limiting variances are difficult to evaluate 
and Baddeley and Gill
(1997) only give explicit results for what they call the sparse
Poisson limit, in which the intensity of a homogeneous
Poisson process tends to 0.

This work studies the estimation of $K$ for a process on $\R$
when the windows are segments of varying lengths.
The fact that the windows are one-dimensional greatly simplifies
the calculation of estimators and permits the explicit derivation of
some of their properties.
The fact that the segment lengths vary provides for an interesting
wrinkle on the approach of Baddeley and Gill (1997).
Notably, simulation results in Section~6 show that the differences
between certain estimators are much greater when the segment
lengths are unequal.

Section~2 describes a cosmological problem that motivated the
present study.
Vanden Berk, et al. (1996) put together a catalog of what are known as 
absorption-line
systems, or absorbers, detected along the lines-of-sight of QSOs
(quasi-stellar objects or quasars).
This catalog, a preliminary version of which
can be obtained from Daniel Vanden Berk (danvb@astro.as.utexas.edu),
provides important evidence
about the large-scale structure of the universe.
To a first approximation, in appropriate units, the locations of these
absorbers along the lines-of-sight can be viewed as multiple realizations
of a stationary point process along segments of varying length.

Section~3 describes the estimators of $K$ used in this paper and gives
explicit expressions for the commonly used rigid motion correction
and isotropic correction estimators
when the observation region is a collection of line segments
of varying lengths.
In addition, Section~3 provides an explicit expression for a modification
to the rigid motion correction advocated in Stein (1993).
The fact that this estimator can be calculated explicitly is in
contrast to the situation in more than one dimension, in which case,
calculating this modified rigid motion correction requires numerous
numerical integrations even for simple regions such as circles and
rectangles.
Finally, following on an idea of Picka (1996), Section~3 introduces
another approach to modifying the rigid motion correction and 
isotropic correction.
When the underlying process is homogeneous Poisson,
Picka's modification of the rigid motion correction has
similar properties to the estimator proposed
in Stein (1993), but theoretical results in Section~5
and simulation results in Section~6 suggest that his approach may
have some advantages and we recommend the adoption of the resulting
estimator for routine use.

When the underlying process is homogeneous Poisson,
Section~4 derives some asymptotic theory for the various estimators
as the number of segments on which the process is observed increases.
As in the case of a single growing observation window studied
in Stein (1993), the modified rigid motion correction 
asymptotically minimizes the variance of the estimator of $K(t)$
among a large class of estimators possessing a type of unbiasedness
property.
Furthermore, if the segments are of equal length, then it is possible
to give explicit comparisons between various estimators.
In particular, the ratio of the asymptotic mean squared
error of the ordinary rigid
motion correction to that
of the modified rigid motion correction equals 1 plus a positive term
proportional to the expected number of events per line segment.
Thus, the benefit of the modification is modest when this expectation
is small, around 1, say, but can be quite substantial when this
expectation is large.

Section~5 considers asymptotic results when the underlying process is
not necessarily homogeneous Poisson, the segments are all of
equal length and the processes on different segments are independent.
In this case, it is essentially trivial to obtain a central limit theorem
for the estimators of $K$ used here.
{}From the general result, it is difficult to make comparisons between
the various estimators.
However, if the process on the different segments are each homogeneous
Poisson but with intensities that vary from segment to segment 
according to some sequence of iid positive random variables,
it is possible to give simple expressions for the asymptotic
variances of the rigid motion correction and the two modifications
of this estimator.
These results show that the modification in Stein (1993)
has strictly smaller asymptotic variance
than the ordinary rigid motion correction.
Furthermore, the modification of Picka (1996) has strictly smaller
asymptotic variance than the modification in Stein (1993) unless the
random intensities have 0 variance, in which case, the two modified
estimators have equal asymptotic variance.

Section~6 reports on the results of a simulation study comparing
the ordinary rigid motion correction and the two modifications
for both Poisson and non-Poisson processes, and equal and unequal
segment lengths.
While there is no theory showing the general superiority
of the modified estimators for non-Poisson processes, the modified
estimators do, for the most part, outperform the unmodified estimator.
The advantage of the modified estimators tend to be larger
when the process is more regular than Poisson,
when the segment lengths are unequal and when $t$ is near the length
of the longest available segment.

Section~7 applies the rigid motion correction and the two modifications
of it described in Section~3 to the estimation of $K$ for the absorber catalog.
In addition, approximate confidence intervals are obtained using
bootstrapping based on viewing the segments as the sampling units.
All three estimates are similar and confirm the finding
in Quashnock and Stein (1999) of clear evidence of clustering
up to at least 50 $h^{-1}$ Mpc.
In addition, the 
confidence intervals based on the modified procedures
produce a slightly stronger case
for clustering of absorbers beyond 100 $h^{-1}$ Mpc.
Whether there is 
clustering of matter at such large scales and for the high
redshifts in the absorber catalog is a critical issue in
modern cosmology, since presently used models for the evolution
of the universe have difficulty explaining such clustering (Steidel, et
al.\ 1998, Jing and Suto 1998).

\beginsection{2.} {The absorber catalog}
The cosmological principle, which states that on large enough spatial
scales, the distribution of matter in the universe is homogeneous
and isotropic, is a central tenet of modern cosmology (Peebles 1993).
In cosmology, it is convenient to measure distances in units of
$h^{-1}$ Mpc, where Mpc, or megaparsec, is 
$3.26\times 10^6$ light years and $h$ is an inexactly known dimensionless
number that is believed to be between 0.5 and 0.75.
As is common in the cosmological literature, in reporting distances
determined from redshifts, we will assume that Hubble's constant,
$H_0$, equals $100\> h\> {\rm km}\> {\rm s}^{-1}\> {\rm Mpc}^{-1}$.
To help calibrate one's thinking about such distances, 1 $h^{-1}$ Mpc is a
typical distance between neighboring galaxies.
It is now generally agreed that galaxies cluster up to scales
of 10--20 $h^{-1}$ Mpc (Davis and Peebles 1983, Loveday, et al.\ 1995).
Furthermore, clustering on such scales can be reproduced
by computer simulations of the evolution of the universe based
on our present understanding of this evolution (see Zhang, et al.\ 1998
and the references therein).
However, there is some evidence of clustering of matter on scales
of up to 100 $h^{-1}$ Mpc (see Quashnock, Vanden Berk and York (1996)
and the references therein) and a few cosmologists have
speculated that clustering may exist at all spatial scales (Coleman
and Pietronero 1992, Sylos Labini, Montuori and Pietronero 1998),
despite the fact that clustering at all scales contradicts both the cosmological
principle and the considerable evidence that supports it
(Peebles 1993, p.\ 20, 45 and 221).
Thus, determining the extent to which clustering of matter is present
is of fundamental importance to modern cosmology.

One way to measure the clustering of matter is through the direct
observation of large numbers of galaxies.
Several galaxy surveys in various regions of the sky
have been done in recent years (Mart\'{\i}nez 1997);
Pons-Border\'{\i}a, et al.\ (1999) describe recent work on estimating
second moment structures of galaxy locations from such
surveys.
The presently ongoing Sloan Digital
Sky Survey will be by far the largest such survey and will contain
roughly $10^8$ galaxies, approximately $10^6$ 
of which will have spectroscopically measured redshifts (Margon 1999).
An object's redshift gives its velocity relative to the Earth,
which, using Hubble's Law, yields its approximate distance from
the Earth.
Galaxy surveys are limited by the fact that galaxies are
difficult to observe directly beyond several hundred $h^{-1}$ Mpc.
QSOs, on the other hand, are extremely bright and focused objects
that can be readily detected at distances of several thousand
$h^{-1}$ Mpc, going back to nearly the beginning of the universe.
Matter that falls on the line-of-sight between the QSO and the Earth
can absorb light from the QSO and thus be detected from the
Earth even though this matter cannot be directly observed.
Certain types of matter absorb light in a characteristic pattern
of frequencies that can be used to identify the matter and, through
the redshift of this absorption pattern, the relative velocity
of this matter to the Earth.
Astronomical objects detected in this way are called absorption-line
systems or absorbers.
As noted by Crotts, Melott, and York (1985), 
catalogs of absorbers provide a means for estimating the clustering
of matter over very large spatial scales.
Vanden Berk et al.\ (1996), Quashnock, Vanden Berk and York (1996)
and Quashnock and Vanden Berk (1998) make use of an extensive catalog
of heavy-element absorption-line systems drawn from the literature
to investigate the clustering of matter at various scales.
York, et al.\ (1991) describe an earlier version of this catalog
and a preliminary version of an updated
catalog is available from Daniel Vanden Berk (danvb@astro.as.utexas.edu).
Here we will use the same absorber catalog as in
Quashnock and Stein (1999), who examined clustering in 352 C~\IV absorbers 
(absorbers detected from the absorption-line patterns of C~{\ninerm IV}, or
triply ionized carbon) along 274 QSO lines-of-sight.
Although the relationship between C~\IV absorbers and galaxies is unclear,
they do appear to track the general spatial patterns of galaxies
(Lanzetta, et al.\ 1995, Quashnock and Vanden Berk 1998), 
and hence provide a plausible means for
assessing the clustering of visible matter on large scales.

Because the universe expands over time and, due to the finite
velocity of light, the more distant an object
the further in the past we observe it, the method used for converting
redshifts into distances from Earth is critical to the analysis of
this catalog.
Redshifts are generally denoted by $z$ and, according to Hubble's
law, an object observed at redshift $z$ is seen at a time
when distances between objects were 
approximately $(1+z)^{-1}$ times their present values.
To correct for the expansion,
here, as in Quashnock and Stein (1999), we use what are called comoving
coordinates, which scale up all distances to what they would be today if
all the matter in the universe moved exactly with the Hubble flow (Peebles
1993).
Thus, in examining the clustering of absorbers in comoving coordinates, we
have removed the most important effects of the universe's expansion.
If one did not make this correction,
the volume density of absorbers would drop approximately like
$(1+z)^3$ as $z$ decreases and we move towards the present.

For various reasons, it is only possible to detect C~\IV
absorbers along a segment of each line-of-sight.
The mean length of these segments in comoving units is 303.3 $h^{-1}$ Mpc,
with a range of 7.5 $h^{-1}$ Mpc to 439.8 $h^{-1}$ Mpc.
For this catalog, the median redshift of the absorbers is about
2.2, with the bulk of absorbers having redshifts from about 1.5 to 3.
Our analysis acts as if clustering is both stationary in time and
homogeneous in space.
We are more accurately examining an average clustering over the range of
redshifts in  the sample at a cosmic epoch corresponding to a 
characteristic redshift of 2.2 (when the universe was about 
1/3 its present scale and about 1/6 its present age).
%
Section 7 provides further discussion of this issue and its possible
influence on our results.

As in Quashnock and Stein (1999), we will act as if the absorber catalog
can be viewed as multiple partial realizations 
of some stationary point process on $\R$ along a series
of segments.
In particular, we will not attempt to use any information about the
physical location of these segments in
three-dimensional space.
Using this simplification, we will then be able to apply the methods
described in the next section to the absorber catalog.

\beginsection{3.}{Methodology}
Suppose $M_1,\ldots,M_p$ are simple, stationary point processes on
$\R$ with a common probability law having intensity $\lambda$ and reduced
second moment function $K$.
We do not necessarily assume that $M_1,\ldots,M_p$ are independent.
For a Borel subset $A$ of $\R$, let $M_j(A)$ be the number of events
of $M_j$ contained in $A$.
If $[0,Q_j]$ is the interval on which we observe $M_j$,
then we can write the observation domain as
$D=\mathop{\cup}\limits_{j=1}^p 
\{[0,Q_j],j\}$, so that $(x,\ell)\in D$ implies
$\ell\in\{1,\ldots,p\}$ and $x\in[0,Q_\ell]$.
Define $N_j=M_j([0,Q_j])$, $N_\sbullet=\sum_{j=1}^p N_j$
and denote the realized value of $N_\sbullet$ by $n$.
For $j=1,\ldots,N_\sbullet$, let $(X_j,L_j)$ be
the random locations of these observed events with realized
values  $(x_j,\ell_j)$ for
$j=1,\ldots,n$.

The basic principle behind all edge-corrected estimators of $K$
described by Ripley (1988)
is to first find an exactly unbiased estimator of $\lambda^2\times
\hbox{volume of observation domain}\times K(t)$ and then to divide
by an estimator of $(\lambda^2\times\hbox{volume})$. 
Here, the volume of the observation domain is $Q_\sbullet=
\sum_{j=1}^p Q_j$.
For a symmetric function $\varphi$ on $D\times D$,
define $T(\varphi)=\sum_{j\ne
k}\varphi\big((X_j,L_j),(X_k,L_k)\big)$.
Then the unbiasedness constraint requires that 
$$
ET(\varphi)= \lambda^2 Q_\sbullet K(t) \eqno(\SNM 1/)
$$
for any reduced moment function $K$.
Estimating $\lambda^2$ by $N_\sbullet(N_\sbullet-1)/Q_\sbullet^2$ yields 
$$
\tilde K(t) = \cases{ \displaystyle{Q_\sbullet T(\varphi)\over N_\sbullet
  (N_\sbullet-1)}
 & if $N_\sbullet>1$,\cr
\noalign{\vskip2pt}
 0 & otherwise\cr}
$$
as a natural estimator
of $K(t)$.
\par
There is an infinite array of functions $\varphi$ satisfying (\Eqn 1/).
Two popular choices are the rigid motion
correction (Ohser and Stoyan 1981) and the isotropic correction
(Ripley 1976).
Asymptotic results in Sections 4 and 5 suggest that
modified versions of the rigid motion correction have good large sample
properties when the underlying process is Poisson, so we focus on
this correction here, although we also give some results for the
isotropic correction for comparison.
It is fairly elementary to prove that the rigid motion 
correction satisfies (\Eqn 1/)
when the observation domain $D$ is a subset of $\R$.
First, for a stationary point process $M$ on $\R$ with intensity
$\lambda$, define the reduced second moment measure ${\cal K}$ by
$\lambda^2 {\cal K}(ds)dx=2E\{M(dx)M(x+ds)\}$, in which case, the reduced
second moment function $K$ is given by $K(t)=\int_{(0,t]}{\cal K}(ds)$.
Denote the indicator function by $1\{\cdot\}$, use $|A|$ to indicate
the Lebesgue measure of the set $A\subset \R$ and $A_s$ to indicate the set $A$
translated by the amount $s$.
The rigid motion correction is given by  
$$
\varphi(x,y)={1\{|x-y|\le t\}|D|\over |D\cap D_{x-y}|}.
$$
We can then write
$$
\eqalignno{
T(\varphi)&=\int_{s\in[-t,0)\cup (0,t]}\int_{x\in \R}M(dx)M(x+ds)
            {1\{x\in D, x+s\in D\}\over |D\cap D_s|}\cr
&=2\int_{s\in (0,t]}\int_{x\in \R}M(dx)M(x+ds)
            {1\{x\in D, x+s\in D\}\over |D\cap D_s|},\cr}
$$
so that 
$$
\eqalignno{
E\{T(\varphi)\} 
&= 2\int_{s\in(0,t]}\int_{x\in \R}{1\over 2}\lambda^2{\cal K}(ds)
  {1\{x\in D, x+s\in D\}\over |D\cap D_s|}\, dx\cr
&=2\int_{s\in(0,t]}{1\over 2}\lambda^2{|D\cap D_s|\over |D\cap
  D_s|}{\cal K}(ds)\cr
&=\lambda^2K(t).\cr}
$$
\par
One way to view the setting where $D$ is a collection of line segments
is to think of these segments as being widely spaced
intervals on $\R$, in which case, we just have a special case
of the treatment in the preceding paragraph.
However, it will be helpful in the subsequent development to think
of $D$ as $\mathop{\cup}\limits_{j=1}^p \{[0,Q_j],j\}$. 
The rigid motion correction can then be defined by taking
$\varphi$ to be
$$
\varphi^R\big((x,k),(y,\ell)\big) = {Q_\sbullet 1\{|x-y|\le t, k=\ell\}
\over \sum_{j=1}^p (Q_j-|x-y|)^+},
$$
where $1\{\cdot\}$ is an indicator function.
To write the isotropic correction in terms of a symmetric
function, let
$$
\varphi^I\big((x,k),(y,\ell)\big)={Q_\sbullet 1\{|x-y|\le t,k=\ell\}
  \{\alpha_\ell(x,y)+\alpha_\ell(y,x)\}\over
   Q_\sbullet-\sum_{j=1}^p\min\{(2|x-y|-Q_j)^+,Q_j\} },
   \eqno(\SNM isodef/)
$$
where $\alpha_\ell(x,y)^{-1}= 1\{x+|y-x|<Q_\ell\}+1\{x-|y-x|>0\}$.
Define
$\tilde K_R(t) = Q_\sbullet T(\varphi^R)/\{N_\bu(N_\bu-1)\}$
and $\tilde K_I(t) = Q_\sbullet T(\varphi^I)/\{N_\bu(N_\bu-1)\}$,
where it is understood that
$\tilde K_R(t)=\tilde K_I(t)=0$ for $N_\bu\le 1$.
We have used Ohser's extension of the isotropic correction
to cover the case $t> {1\over 2}\min(Q_1,\ldots,Q_p)$ (Ohser 1983).
As Ripley (1988, p.~32) notes, this extension is generally not
of much practical value when there is a single contiguous observation
window.
However, when there are multiple windows of various sizes, the extension
is critical.
For the absorber catalog, for example,
one is certainly interested in estimating
$K$ at distances greater than $3.75$~$h^{-1}$ Mpc, the value of
${1\over 2}\min(Q_1,\ldots,Q_p)$ in the catalog.
\par
Note that $\varphi^I
\big((x,k),(y,\ell)\big)=\varphi^R\big((x,k),(y,\ell)\big)=0$
if $k\ne \ell$, which just says that pairs of observations on different
segments do not contribute to the estimate of $K(t)$.
Since we have made no assumption about the joint distribution
of $M_1,\ldots,M_p$, for (\Eqn 1/) to be valid, it is necessary to assume
$\varphi\big((x,k),(y,\ell)\big)=0$
whenever $k\ne \ell$.
Thus, throughout this work, we will only consider $\varphi$ satisfying
\item{(A)}
$\varphi\big((x,k),(y,\ell)\big)=0$ for $k\ne \ell$.
\par
We next show how to apply to the present setting
the method developed in Stein (1993) for improving
upon any estimator of $K$ of the form $Q_\sbullet
T(\varphi)/\{N_\bu(N_\bu-1)\}$ with $\varphi$ satisfying (\Eqn 1/).
Suppose $(X,L)$ is uniformly distributed on $D$ in the sense that
$P(L=\ell)=Q_\ell/Q_\sbullet$ and the density of $X$ given
$L=\ell$ is uniform on $[0,Q_\ell]$.
Then $M_1,\ldots,M_p$ stationary with common distribution imply
that for any real-valued function $g$ for which $E|g(X,L)|<\infty$,
$E\sum_{j=1}^{N_\bu}g(X_j,L_j)=\lambda Q_\sbullet Eg(X,L)$,
so that $\sum_{j=1}^{N_\bu}\{g(X_j,L_j)-Eg(X,L)\}$ 
is an unbiased estimator of 0.
The idea in Stein (1993) is to choose $g$ to minimize
$$
\var_n\biggl[ T(\varphi) - \sum_{j=1}^n\{g(X_j,L_j)-Eg(X,L)\}\biggr],
$$
where $\var_n$ means to compute the 
variance under binomial sampling:
$N_\bu=n$ is fixed and, for $j=1,\ldots,n$,
 $(X_j,L_j)$ are independent
and all have the same distribution as $(X,L)$.
Proposition~1 in Stein (1993) shows that for $n\ge 1$ and $(y,m)\in D$,
a minimizing $g$ is $2(n-1)h(y,m;\varphi)/Q_\bu$, where
$h(y,m;\varphi) = \sum_{\ell=1}^p\int_0^{Q_\ell}\varphi\big((x,l),(y,m)\big)
dx$.  
Under (A), 
$h(y,m;\varphi) = \int_0^{Q_m}
\varphi\big((x,m),(y,m)\big)dx$.

Now define
$$
T^*(\varphi)=T(\varphi)-{2(N_\bu-1)\over Q_\bu}
  \sum_{j=1}^{N_\bu}\left\{h(X_j,L_j;\varphi)
-Eh(X,L;\varphi)\right\}.
$$
Note that if $\varphi$ satisfies (\Eqn 1/),
$Eh(X,L;\varphi)=2t$.
Under binomial sampling, we always have $\var_n\{T^*(\varphi)\}\allowbreak\le
\var_n\{T(\varphi)\}$.
This suggests that the estimator  
$\hat K(t) = Q_\sbullet T^*(\varphi)/\{N_\bu(N_\bu-1)\}$ for $N_\bu>1$
and 0 otherwise may be preferred over $\tilde K(t)$.
As with the unmodified estimators, $\hat K_R(t)$ indicates that
$\varphi=\varphi^R$ and $\hat K_I(t)$ indicates that $\varphi=\varphi^I$.
\par
Picka (1996) suggests another approach to modifying estimates of second moment
measures.
He considered random sets for which the probability of any fixed point being
in the random set is positive, but his approach
can also be applied to point processes, for which this probability is 0.
For point processes, his idea corresponds to using an estimator of
$\lambda Q_\sbullet$ other than $N_\bu$ in $\tilde K$.
For any real-valued function $c$ on $D$ satisfying $\sum_{\ell=1}^p
\int_0^{Q_\ell} c(x,\ell)dx=Q_\sbullet$, $\hat\lambda_c=Q_\sbullet^{-1}
\sum_{j=1}^{N_\bu} c(X_j,L_j)$ is an unbiased estimator of $\lambda$.
Let us consider estimators of $K(t)$ of the form $Q_\sbullet T(\varphi)/
\{\hat\lambda_c Q_\sbullet(\hat\lambda_c Q_\sbullet - 1)\}$.
It is not generally possible to calculate the exact variance
of such estimators under binomial sampling.
However, for $Q_\sbullet$ sufficiently large, $\hat\lambda_c-\lambda$
and $Q_\sbullet^{-1}T(\varphi)-\lambda^2 K(t)$ should be small in probability,
which suggests using a first-order Taylor series approximation to
obtain
$$
{Q_\sbullet T(\varphi)\over \hat\lambda_c Q_\sbullet(\hat\lambda_c Q_\sbullet
- 1)}\approx {1\over \lambda^2 Q_\sbullet} T(\varphi) - {2K(t)\over \lambda}
(\hat\lambda_c-\lambda).\eqno(\SNM 1.2/)
$$
For a given $\varphi$ and subject to $c$ satisfying the unbiasedness
constraint, now consider minimizing the variance
of the right side of (\Eqn 1.2/) when $M_1,\ldots,M_p$
are iid Poisson processes with intensity $\lambda$.
It is a straightforward variational problem to show that a minimizing $c$ 
is given by $c(x,\ell;\varphi)=h(x,\ell;\varphi)/(2t)$.
Define
$$
\Kvee(t) = {Q_\sbullet T(\varphi)\over \sum_{j=1}^{N_\bu}
c(X_j,L_j;\varphi)\big\{\sum_{j=1}^{N_\bu} c(X_j,L_j;\varphi) - 1
\big\}}
$$
for $N_\bu>1$ and $\Kvee(t)=0$ otherwise.
As with $\tilde K$ and $\hat K$, subscripts $R$ or $I$ on $\Kvee$
indicate that $\varphi=\varphi^R$ or $\varphi=\varphi^I$.
\par
When $M_1,\ldots,M_p$
are iid Poisson processes,
$\hat K(t)$ and $\Kvee(t)$ should behave similarly.
To see this, first use Taylor series to obtain
$$
\hat K(t)\approx{1\over \lambda^2 Q_\sbullet} T(\varphi) - {2\over\lambda
Q_\bu} \sum_{j=1}^{N_\bu} h(X_j,L_j;\varphi)+2\{2t-K(t)\}{N_\bu\over \lambda
Q_\bu} + 2K(t).
$$
{}From this approximation and (\Eqn 1.2/),
when $K(t)=2t$, both $\hat K$ and $\breve K$ are approximately
$$
{1\over \lambda^2 Q_\sbullet} T(\varphi) - {2\over\lambda Q_\bu}
\sum_{j=1}^{N_\bu} h(X_j,L_j;\varphi) + 4t.
$$
Thus, for $Q_\sbullet$ large, the two estimators will be similar
when $M_1,\ldots,M_p$
are iid Poisson processes,
but they are not necessarily similar otherwise.
\par
Even for simple regions in two or more dimensions, calculating
$h(\cdot;\varphi)$ requires numerical integrations.
However, when the observation region is 
$D=\mathop{\cup}\limits_{j=1}^p 
\{[0,Q_j],j\}$, then it is possible to give
an explicit expression for $h(x,\ell;\varphi^R)$ for 
$(x,\ell)\in D$.
For convenience, we will assume that the $Q_j$s have been arranged in
increasing order.
For $r<Q_p$, define $j(r)=\min_{1\le j\le p}\{j: Q_j\ge r\}$ and
let $U(r)=\sum_{j=1}^p (Q_j-r)^+$.
For $j=1,\ldots,p$, let $U_j=U(Q_j)$
and set $Q_0=0$ so that $U_0=Q_\sbullet$.
Furthermore, define
$$
\kappa(x,t) =  \sum_{j=1}^{j(x\land t)-1} {1\over p-j+1}\log\left(
              {U_{j-1}\over U_j}\right)
 + {1\over p-j(x\land t)+1}\log\left\{ {U_{j(x\land t)-1}\over
    U(x\land t)}\right\},
$$
where a sum whose upper limit is less than its lower limit is defined
to be 0 and $x\land t$ is the minimum of $x$ and $t$.
Then 
$$
Q_\bu^{-1}h(x,\ell;\varphi^R) = \kappa(x,t) + \kappa(Q_\ell-x,t)\eqno(\SNM 2/)
$$ 
(see the appendix).
If the segment lengths are all equal, $\kappa(x,t)=p^{-1}\log
[Q/\{Q-(x\land t)\}]$.
\par
It is also possible to evaluate $h(x,\ell;\varphi^I)$ explicitly,
but the resulting expression is rather cumbersome.
If $t<{1\over 2}\min(Q_1,\cdots, Q_p)$,
then the denominator in the definition of $\varphi^I$ in
(\Eqn isodef/) equals $Q_\bu$ whenever $|x-y|\le t$, which
greatly simplifies matters.
In this case, it is possible to show that
$$
h(x,\ell;\varphi^I)= t + (x\land t) + \{(Q_\ell-x)\land t\} -{1\over 2}
\left({x\over 2}\land t\right)-{1\over 2}\left({Q_\ell-x\over 2}\land t
\right).
$$
A second special case yielding a simple result is when $Q_1=\cdots=Q_p
=Q$.
When $t<{1\over 2}Q$, the preceding expression for $h$ applies
and for $t\ge {1\over
2}Q$,
$$
 h(x,\ell;\varphi^I)= 
 {3Q\over 4} + \{x\land(Q-x)\} + Q\log\left[{{1\over 2} Q
\over \{x\land(Q-x)\}\lor(Q-t)}\right],
$$
where $x\lor y$ is the maximum of $x$ and $y$.

There is a considerable literature in astrophysical journals
on estimating second order characteristics of galaxy locations
based on galaxy surveys in large, contiguous regions of the
sky.
Mart\'{\i}nez (1997) and Stoyan and Stoyan (2000) provide two
recent reviews of this work.
Astrophysicists have generally focused on estimating the pair
correlation function, which is, after a normalization, just
the derivative of the $K$ function.
For example, for a stationary point process
$M$ on $\R$, assuming $K$ is differentiable,
the pair correlation function is ${1\over 2}K'$.
Similar to $\hat K$ here,
Landy and Szalay (1993) make use of unbiased estimators of
0 to modify estimators of second order characteristics.
Moreover, similar to $\breve K$,
Hamilton (1993) describes estimators of the pair
correlation function of the form $T(\phi)/{\hat\lambda}^2$
in which $\lambda^2$ is estimated by something other than
the obvious estimator.
We prefer to estimate $K$ rather than the pair correlation function
because it separates the problem of handling edge effects
from that of density estimation and the consequent smoothing problem.
If one wants to estimate the pair correlation function, we
recommend first computing an appropriately edge-corrected
estimate of $K$ and then differentiating a smoothed version
of this estimate.

\beginsection{4.}{Asymptotic theory when the truth is Poisson}
There are a number of ways one might take limits to study the properties
of the estimators proposed in the previous section.
One possibility would be to fix $p$ and let the $Q_j$s tend
to $\infty$.
In this approach, the fraction of the observation
region within a fixed distance of an endpoint of a segment tends
to 0 and, as in Ripley (1988) and Stein (1993), the 
variance of all reasonable estimators of $K(t)$ 
for fixed $t$ have the same first-order
asymptotic behavior under binomial sampling.
However, for the absorber catalog, in which $p=274$ and
the number of absorbers per line is $1.28$, a more relevant choice
is to uniformly bound the $Q_j$s and let $p\to\infty$.
This limiting approach keeps the fraction of the observation
region within a fixed distance of an endpoint of a segment bounded
away from 0 with the result that the differences between various
estimators under binomial sampling show up in the leading terms
for the asymptotic variance.
Hansen, Gill and Baddeley (1996) and
Baddeley and Gill (1997) take a similar asymptotic approach
for studying estimators of properties
of spatial point processes based on observing the process
in an increasing number of identical and distantly spaced
windows.

We now consider adapting the 
asymptotic results in Ripley (1988) and Stein (1993)
to the present setting.
First, we give exact expressions for the variance under binomial
sampling of both $\tilde K(t)$ and $\hat K(t)$.
Following Ripley (1988),
for a symmetric function $\varphi$ on $D\times D$ satisfying (A),
define
$$
S(\varphi)=\sum_{j=1}^p \int_0^{Q_j}\! \int_0^{Q_j}
\varphi\big((x,j),(y,j)\big)dx\, dy,
$$
$$
S_1(\varphi)=
\sum_{j=1}^p \int_0^{Q_j}\left\{ \int_0^{Q_j}
\varphi\big((x,j),(y,j)\big)dx\right\}^2\! dy,
$$
and
$$
S_2(\varphi)=\sum_{j=1}^p \int_0^{Q_j}\! \int_0^{Q_j}
\varphi\big((x,j),(y,j)\big)^2 dx\, dy.
$$
Under (A) (Ripley 1988),
$$
\var_n\{T(\varphi)\}={2n(n-1)\over Q_\sbullet^2}\left\{ 
S_2(\varphi) + {2n-4\over Q_\sbullet}S_1(\varphi) - {2n-3\over
Q_\sbullet^2}S(\varphi)^2\right\}\eqno(\SNM varT/)
$$
and (Stein 1993)
$$
\var_n\{T^*(\varphi)\}={2n(n-1)\over Q_\sbullet^2}\left\{
S_2(\varphi) - {2\over Q_\sbullet}S_1(\varphi) + {1\over
Q_\sbullet^2}S(\varphi)^2\right\}.\eqno(\SNM 3/)
$$
\par
We now want to study what happens as $p\to\infty$.
Suppose $Q_1,Q_2,\ldots$ is a sequence of positive numbers
and the subscript $p$ is used to indicate the dependence of a 
term on the number of segments observed, so that
$D_p=\mathop{\cup}\limits_{j=1}^p
\{[0,Q_j],j\}$, $Q_{\sbullet p}=\sum_{j=1}^p Q_j$ and $N_{\bu p}$ is the total
number of events on $D_p$.
Suppose $\{\varphi_p\}$ is a sequence of functions for which the
domain of $\varphi_p$ is $D_p\times D_p$ and $\varphi_p$ is symmetric
for all $p$.
In addition to $\varphi_p$ satisfying (A) for all $p$,
we will assume the following regularity conditions:
\item{(B)} The $\varphi_p$s are uniformly bounded;
\item{(C)} For each $p$, $\varphi_p$ satisfies the unbiasedness
constraint in (\Eqn 1/);
\item{(D)} The $Q_j$s are bounded away from 0 and $\infty$.
\par\noindent
Under (A)--(D), we have $S(\varphi_p)=2tQ_{\sbullet p}=O(p)$,
$S_1(\varphi_p)=O(p)$ and $S_2(\varphi_p)=O(p)$ but is not $o(p)$.
It follows that as $p\to\infty$,
$$
S_2(\varphi_p) - {2\over Q_{\sbullet p}}S_1(\varphi_p) + {1\over
Q_{\sbullet p}^2}S(\varphi_p)^2=
S_2(\varphi_p)\big\{1+O\big(p^{-1}\big)\big\}.\eqno(\SNM 4/)
$$
Comparing (\Eqn 3/) and (\Eqn 4/) suggests that minimizing $S_2(\varphi_p)$
subject to (A)--(D) is nearly the same as minimizing
$\var_n\{T^*(\varphi_p)\}$.
Stein (1993) shows that subject to (C), 
the rigid motion correction gives a minimizer of $S_2(\varphi_p)$.
The appendix gives an explicit expression for $S_2(\varphi^R)$ in
terms of elementary functions.
\par
We next obtain an analog to Proposition~2 in
Stein (1993), which demonstrates the asymptotic optimality under
the Poisson model for $\hat K_R$ among a certain class of estimators
as the dimensions of a single observation window increase. 
For a sequence of functions $\{\varphi_p\}$ on $D_p\times D_p$
and a sequence of functions $\{g_p\}$ on $D_p\times \{0,1,\ldots\}$,
define the statistic $\Theta(\varphi_p,g_p)$ by
$$
\Theta(\varphi_p,g_p)={Q_{\sbullet p}\over N_{\bu p}(N_{\bu p}-1)}
\left[ T(\varphi_p)-\sum_{j=1}^{N_{\bu p}}\left\{g_p\big((X_j,L_j),N_{\bu
p}\big)-{1\over Q_{\sbullet p}}\sum_{\ell=1}^p \int_0^{Q_\ell}
g_p\big((x,\ell),N_{\bu p}\big)dx\right\}\right]
$$
if $N_{\bu p}>1$ and 0 otherwise.
Write $E_\lambda$ to indicate expectations assuming $M_1,M_2,\ldots$
are independent Poisson processes with constant intensity $\lambda$
independent of $p$.
All ensuing asymptotic results
in the rest of this section involve expectations over the Poisson model
and can be proven by first conditioning on $N_{\bu p}$,
using the fact that under this model,
the conditional distribution of the observed events on $D_p$ 
follows binomial sampling, and finally, by averaging over the
distribution of $N_{\bu p}$, which follows a Poisson distribution with mean 
$\lambda Q_{\sbullet p}$.
\Proposition{1}. Suppose $\{\varphi_p\}$ satisfies (A)--(C),
$E_\lambda\Biggl\{\displaystyle\sum_{j=1}^{N_{\bu p}}
\big|g_p\big((X_j,L_j),N_{\bu
p}\big)\big|\Biggr\}<\infty$ for all $p$, the $Q_j$s satisfy (D)
and $p^{-1}\sum(Q_j-t)^+$ is bounded away
from 0 as $p\to\infty$.  Then
$$
p^2\left[E_\lambda\left\{\hat K_R(t)
  -2t\right\}^2-E_\lambda\left\{
 \Theta(\varphi_p,g_p) -2t\right\}^2\right]
$$
is bounded from above as $p\to\infty$.
\TheoremEnd
\par\noindent
The assumption that $p^{-1}\sum(Q_j-t)^+$ is bounded away
from 0 as $p\to\infty$ guarantees that $\{\varphi^R_p\}$ satisfies (B).
Since, under the conditions of Proposition 1,
$E_\lambda\big\{{\hat K}_R(t)-2t\big\}^2=O(p^{-1})$ as $p\to\infty$,
this result says that when the underlying
processes are independent Poisson with equal
intensity, ${\hat K}_R$ asymptotically minimizes
the mean squared error among all sequences of estimators of the
form considered in the proposition.
\par
Let us now make some comparisons of the asymptotic mean squared errors
of some estimators of $K(t)$ under the Poisson model
when all $Q_j$s equal $Q$ and $s=t/Q$.
{}From (\Eqn 3/), we get $E_\lambda\left\{\hat K(t)-2t\right\}^2\sim
{2\over \lambda^2 p^2 Q^2}S_2(\varphi_p)$.
Thus, (17) 
in the appendix implies
$$
E_\lambda\left\{\hat K_R(t)
  -2t\right\}^2\sim -{4\over \lambda^2 p}\log(1-s)\eqno(\SNM 4.8/)
$$
and (20) 
in the appendix implies
$$
E_\lambda\left\{\hat K_I(t)
  -2t\right\}^2\sim   
{4\over \lambda^2 p}\times \cases
{  s +{3\over 4}s^2 & if $0<s\le {1\over 3}$,\cr
 {1\over 12}+ {1\over 2}s+ {3\over
  2}s^2& if ${1\over 3}\le s\le {1\over 2}$ and\cr
{17\over 24} - \log 2 -\log(1-s)&
   if ${1\over 2}\le s<1$.\cr}\eqno(\SNM 5/)
$$
{}From Proposition~1, the right side of (\Eqn 5/) must be at
least as large as the right side of (\Eqn 4.8/) for all $s\in(0,1)$.
In fact,
it is a straightforward exercise to show analytically that
the right side of (\Eqn 5/) is strictly greater than the right
side of (\Eqn 4.8/) for all $s\in(0,1)$.
Thus, as $p\to\infty$, the modified rigid motion estimator $\hat K_R$ performs
nonnegligibly better than either the ordinary 
or modified isotropic estimator 
for any $t\in (0,Q)$ under the Poisson model, although the improvement
over the modified isotropic estimator is minor.               
Figure~1 shows the ratio of the asymptotic variances for
$\hat K_I(t)$ and $\hat K_R(t)$ under the Poisson model, which
reaches a maximum of approximately $1.032$ near $t=0.247Q$.
The asymptotic results in (\Eqn 4.8/) and (\Eqn 5/) are unchanged
if $\Kvee_R$ and $\Kvee_I$ replace $\hat K_R$ and $\hat K_I$.
\par
We next compare the modified and unmodified rigid motion estimators
as $p\to\infty$ when all $Q_j$s equal $Q$.
{}From (\Eqn varT/),
$$
E_\lambda\left\{\tilde K(t) -2t\right\}^2\sim {2\over \lambda^2 p^2 Q^2}
S_2(\varphi^p) + {4\over \lambda p^2 Q^2} S_1(\varphi^p) - {16t^2\over
\lambda p Q}.
$$
Using (17) and (18) 
in the appendix then yields  
$$
E_\lambda\left\{\tilde K_R(t)
  -2t\right\}^2\sim  {4\over \lambda^2 p}\left[-\log(1-s)
+4\lambda Q\{\gamma(s)-s^2\}\right],\eqno(\SNM unmod/)
$$
where
$$
\gamma(s)=
 {1\over 4}\int_0^1\left[\int_0^1 {1\{|x-y|\le s\}
  \over 1-|x-y|}dy\right]^2 dx.\eqno(\SNM gamma.def/)
$$
Equation (19) 
in the appendix gives a more explicit expression for $\gamma$.
Note that
$$
\gamma(s)-s^2= {1\over 4}\int_0^1\left[\int_0^1 {1\{|x-y|\le s\}
  \over 1-|x-y|}dy-2s\right]^2 dx,
$$
which is strictly positive for all $s\in (0,1]$.
\par
Comparing (\Eqn 4.8/) and (\Eqn unmod/) shows that,
in terms of mean squared error,
the asymptotic relative advantage of either modified rigid motion
estimator over the unmodified
rigid motion estimator is proportional to $\lambda Q$, the expected
number of events per segment.
Figure~2 plots $4\{\gamma(s)-s^2\}/\{-\log(1-s)\}$, which is less
than 0.124 for all $s\in (0,1)$ and is less than 0.061 for all $s<0.9$.
Thus, at least for equal $Q_j$s, we should not
expect a large improvement under the Poisson model due to the
modifications when there are only 1.28 events per segment as in the
absorber catalog.
Simulation results in Section~6 show that larger improvements can occur
with unequal $Q_j$s.
\par
\beginsection {5.} {Some asymptotic theory for non-Poisson processes}
There is a decided lack of asymptotic theory that permits useful comparisons
of estimators of $K$ when the underlying process is not Poisson.
Stein (1995) derives results showing the advantage of estimators like
$\hat K$ over those like $\tilde K$, but the asymptotic approach
taken there requires that the distance $t$ at which one is estimating
$K$ be large compared to the distances at which the underlying process
shows nontrivial dependence.
When the observation window is made up of many
segments, especially if the $Q_j$s are equal
and the $M_j$s are independent, it appears feasible to develop
some useful asymptotic results for non-Poisson processes.
This section describes some general asymptotic results for the
estimators $\tilde K$, $\hat K$ and $\breve K$ described in Section~3.
These results are used to demonstrate that if $M_1,M_2,\ldots$
are, conditional on $\Lambda_1,\Lambda_2,\ldots$, independent Poisson
processes with $M_j$ having intensity $\Lambda_j$, where the
$\Lambda_j$s are iid positive
random variables, then as $p\to\infty$, $\Kvee_R(t)$ is superior
to $\hat K_R(t)$, which is in turn superior to $\tilde K_R(t)$.
\par
Suppose $M_1,M_2,\ldots$ are iid simple, stationary point processes
on $\R$ with intensity $\lambda$ and reduced second moment
function $K$.
Assume $Q=Q_1=Q_2=\cdots$ and
let $X_{1j},\ldots,X_{N_j j}$ be the locations
of the $N_j$ events from $M_j$ on $(0,Q)$.
For a bounded, symmetric function $\varphi$ on $(0,Q)\times (0,Q)$, define
$\Phi_j=\sum_{k\ne \ell}\varphi(X_{kj},X_{\ell j})$. 
Analogous to (\Eqn 1/), suppose $E\Phi_j=\lambda^2Q K(t)$ for any
reduced second moment function $K$ for the $M_j$s.
Define $G_j=(2t)^{-1}\sum_{k=1}^{N_j}\int_0^Q \varphi(X_{kj},y)dy$,
so that $EG_j=\lambda Q$.
Using these definitions, the estimators described in Section~3 are
given by
$$\eqalignno{
\tilde K(t) &= {pQ\sum_{j=1}^p\Phi_j\over \sum_{j=1}^p N_j\left(\sum_{j=1}^p
N_j-1\right)},\cr
\hat K(t) &= \tilde K(t) - {4t\sum_{j=1}^p G_j\over \sum_{j=1}^p N_j} + 4t\cr
\noalign{\hbox{and}}
\Kvee(t) &= {pQ\sum_{j=1}^p\Phi_j\over \sum_{j=1}^p G_j\left(\sum_{j=1}^p
G_j-1\right)}.\cr}
$$
Furthermore, since $\{N_j,\Phi_j,G_j\}_{j=1}^\infty$ is an iid trivariate
sequence, we can readily derive the limiting distribution of these
estimators.
Specifically, if $E(N_1^4)<\infty$, then $\Phi_1$ and $G_1$ have finite
second moments, so as $p\to\infty$,
$$
p^{1/2}\pmatrix{{1\over p}\sum_{j=1}^p N_j-\lambda Q\cr
		\noalign{\vskip 2pt}
		{1\over p}\sum_{j=1}^p \Phi_j-\lambda^2QK(t)\cr
		\noalign{\vskip 2pt}
		{1\over p}\sum_{j=1}^p G_j-\lambda Q\cr}
\Larrow N(0,\Sigma),
$$
where $\Larrow$ indicates convergence in distribution and $\Sigma$ is
the $3\times 3$ covariance matrix of $(N_1,\Phi_1,G_1)$.
Using first-order Taylor series, we get $\lambda Q p^{1/2}\{\tilde K(t) -
K(t)\}\Larrow N(0,\tilde V)$, $\lambda Q p^{1/2}\{\tilde K(t) - K(t)\}
\Larrow N(0,\hat V)$ and $\lambda Q p^{1/2}\{\Kvee(t) - K(t)\}\Larrow N(0,
\Vvee)$, where
$$
\tilde V = 4K(t)^2\var(N_1)
+{1\over \lambda^2}\var(\Phi_1)-{4K(t)\over \lambda}\cov(N_1,\Phi_1),
\eqno(\SNM CLT1/)
$$
$$
\eqalign{
\hat V &= 4\{K(t)-2t\}^2\var(N_1)
 +{1\over \lambda^2}\var(\Phi_1)+16t^2\var(G_1)-{4\{K(t)-2t\}\over \lambda}
 \cov(N_1,\Phi_1)\cr
&\qquad - {8t\over \lambda}\cov(\Phi_1,G_1)+16\{K(t)-2t\}\cov(N_1,G_1)\cr}
\eqno(\SNM CLT2/)
$$
and
$$
\Vvee = 4K(t)^2\var(G_1)
+{1\over \lambda^2}\var(\Phi_1)-{4K(t)\over \lambda}\cov(\Phi_1,G_1).
\eqno(\SNM CLT3/)
$$
As expected, $\Vvee = \hat V$ when $K(t)=2t$.
\par
To calculate the limiting behavior of these estimators
for any given $\varphi$, $Q$ and law of $M_1$, we only have to compute
the covariance matrix $\Sigma$ and plug the results into (\Eqn CLT1/)--(\Eqn
CLT3/).
In some limited cases this computation can be done analytically or
more often by numerical integration; otherwise,
$\Sigma$ is easily approximated by simulation whenever $M_1$ can be readily 
simulated.
\par
We now consider a simple setting in which $\Sigma$ can be explicitly
derived.
Suppose $M_1,M_2,\ldots$
are, conditional on $\Lambda_1,\Lambda_2,\ldots$, independent Poisson
processes with $M_j$ having intensity $\Lambda_j$, where the
$\Lambda_j$s are iid positive
random variables.
Such a model could serve as an approximation for a Cox process 
(Daley and Vere-Jones 1988, Section 8.5) observed
over widely spaced segments where the random intensity function 
$\Lambda(\cdot)$ of the process has little
variation over distances of length $Q$ but the segments are sufficiently
spaced so that the behavior of $\Lambda(\cdot)$ in different segments is 
essentially independent.
\par
Next, suppose $\varphi(x,y)=Q1\{|x-y|\le t\}/(Q-|x-y|)$, 
so that we are using the rigid motion estimator.
In this case, the elements of $\Sigma$
can be readily calculated in terms of the moments of $\Lambda_1$.
Writing $m_j$ for $E(\Lambda_1^j)$, we have $\lambda=m_1$, 
$K(t)=2tm_2/m_1^2$,
$$\eqalignno{
\var(N_1)&= Qm_1+Q^2(m_2-m_1^2),\cr
\var(\Phi_1)&= 16Q^3\gamma\left({t\over Q}\right)m_3-4Q^2\log\left(1-{t\over
 Q}\right)m_2+4t^2Q^2(m_4-m_2^2), \cr
\var(G_1)&= {Q^3\over t^2}\gamma\left({t\over Q}\right)m_1+Q^2(m_2-m_1^2),\cr
\cov(N_1,\Phi_1)&= 4tQm_2+2tQ^2(m_3-m_1m_2),\cr
\cov(N_1,G_1)&=Qm_1+Q^2(m_2-m_1^2)\cr
\noalign{\hbox{and}}
\cov(\Phi_1,G_1)&= {4Q^3\over t}\gamma\left({t\over Q}\right)m_2
 +2tQ^2(m_3-m_1m_2).\cr}
$$
Each of these results can be obtained by conditioning on $\Lambda_1$.
For example,
$$\eqalignno{
\var(\Phi_1) &= E\{\var(\Phi_1\mid \Lambda_1)\}+\var\{E(\Phi_1\mid
 \Lambda_1)\}\cr
&= E\left[4\Lambda_1^3
 \int_0^Q\left\{\int_0^Q \varphi(x,y)dy\right\}^2\! dx
  +2\Lambda_1^2\int_0^Q\int_0^Q\varphi(x,y)^2dx\,dy\right]
  +\var(2t\Lambda_1^2Q)\cr
&= 16Q^3\gamma\left({t\over Q}\right)m_3-4Q^2\log\left(1-{t\over
 Q}\right)m_2+4t^2Q^2(m_4-m_2^2), \cr}
$$
where the second step follows from (10) in Ripley (1988, p.~30) and the
last step uses (17) 
and (18) 
in the appendix.
\par
Plugging these results into (\Eqn CLT1/)--(\Eqn CLT3/) yields
$$
\eqalignno{
\tilde V_R &= {1\over m_1^2}\var(\Phi_1) -16t^2Q{m_2^2\over m_1^3}
 +16t^2Q^2 {m_2(m_2^2-m_1m_3) \over m_1^4},\cr
\hat V_R &= {1\over m_1^2}\var(\Phi_1) -16t^2Q{(m_2-m_1^2)^2\over m_1^3}
 -16Q^3\gamma\left({t\over Q}\right)\left({2m_2\over m_1}-m_1\right)
 +16t^2Q^2 {m_2(m_2^2-m_1m_3)\over m_1^4}\cr
\noalign{\hbox{and}}
\Vvee_R &= {1\over m_1^2}\var(\Phi_1) -16Q^3\gamma\left({t\over Q}\right)
 {m_2^2\over m_1^3} +16t^2Q^2 {m_2(m_2^2-m_1m_3)\over m_1^4},\cr}
$$
where the subscript $R$ indicates that the asymptotic variance
is for the appropriate version of the rigid motion estimator.
Thus,
$$
\tilde V_R -\hat V_R= 16Q^3\left({2m_2\over m_1}-m_1\right)
 \left\{\gamma\left(t\over Q\right)-{t^2\over Q^2}\right\},\eqno(\SNM V1/)
$$
which is positive on $(0,1)$ 
since $\gamma(s)-s^2>0$ for $s\in(0,1)$
and $m_2\ge m_1^2$.
Furthermore, 
$$
\hat V_R-\Vvee_R= 16Q^3{(m_2-m_1^2)^2\over m_1^3}\left\{\gamma\left({t\over
Q}\right)-{t^2\over Q^2}\right\},\eqno(\SNM V2/)
$$
which is positive on $(0,1)$ whenever $m_2>m_1^2$.
Thus, $\hat V_R>\Vvee_R$ unless $\var\Lambda_1=0$, in which case,
$m_2=m_1^2$ and $\hat V_R=\Vvee_R$.
\par
The arguments in this section largely carry over to estimators 
for the reduced second moment function of iid point processes
on $\R^d$ observed over 
$\mathop{\cup}\limits_{j=1}^p 
\{A,j\}$ for some $A\subset \R^d$.
In particular, (\Eqn CLT1/)--(\Eqn CLT3/) still hold if, at the appropriate
places, $2t$ is replaced by $\mu_dt^d$,
the volume of a ball of radius $t$ in $\R^d$.
Furthermore, the comparisons between $\tilde V_R$, $\hat V_R$ and $\Vvee_R$ in
(\Eqn V1/) and (\Eqn V2/) still hold after replacing $\gamma(t/Q)-t^2/Q^2$
by $\int_A\big\{\int_A \varphi(x,y) dy-\mu_d t^d\big\}^2dx$.
\par
\beginsection {6.} {Simulation study}
The asymptotic results in the preceding two sections provide only limited
information about the relative advantages of the various estimators, especially
for non-Poisson processes or unequal $Q_j$s.
Because the estimators $\tilde K_R$, $\hat K_R$ and $\Kvee_R$ can all be
explicitly calculated, it is fairly straightforward to study the behavior
of these estimators via simulation.
This section reports some results from a simulation study that considers
equal and unequal $Q_j$s and three models for the law of the point processes.
For the unequal segment length case, $p=50$ and $Q_j=0.1j$ for $j=1,\ldots,p$
and for the equal segment length case, $p=50$ and each $Q_j=2.55$, so that
$Q_\bu=127.5$ in both cases.
The three processes reported on here are all stationary renewal processes;
that is, the waiting times between consecutive events are iid random variables.
In each case, the intensity of the process is 1, so that $EN_\bu=127.5$
in all simulations.
Stationary renewal processes are straightforward to simulate on an
interval $[0,Q]$.
If $F$ is the cdf (cumulative distribution function) for the waiting
times and $\mu<\infty$ is the mean waiting time, then to obtain
a stationary process on $[0,\infty)$, use $\mu^{-1}\int_0^x
\{1-F(y)\}dy$ for the cdf of the time of the first event after 0
(Daley and Vere-Jones 1988, p.~107).
Simulate a random variable from this distribution; if it is greater than
$Q$ then one is done and there are no events in $[0,Q]$ for this realization
of the process.
If not, simulate random waiting times with cdf $F$ until
one gets the first event after $Q$ and use the preceding events as
the realization of the process on $[0,Q]$.
Here, we consider waiting time densities $f$ that are exponential with
mean 1 (in which case the $M_j$s are Poisson processes), 
$f(x)=4xe^{-2x}$ for $x>0$
(a gamma density with parameters 2 and ${1\over 2}$) 
and $f(x)=24/(2+x)^4$ for $x>0$.
Figure~3 plots $K(t)-2t$ for renewal processes
with the last two waiting time densities,
which shows that the first of these corresponds to a process more
regular than the Poisson and the second is more clumped than the Poisson.
For the gamma waiting times, it is possible to show that for $x\ne 0$,
$P\{M_1(dx)=1\mid M_1(\{0\})=1\}=1-e^{-4x}$ and hence that $K(t)=2t-{1\over 2}
(1-e^{-4t})$.
For the third waiting time density, we cannot give an analytic expression
for $K(t)$, although Theorem~1 in Feller (1971, p.~366) implies that
$K(t)-2t\to 2$ as $t\to\infty$.
The values for $K(t)$ in Figure~3 for this process were obtained by simulation.
Since the mean waiting times are all equal, the variances of the waiting
times provide another measure of clumpiness with larger variances
corresponding to a clumpier process.
For the exponential
waiting times, the variance is 1, for the gamma case, the variance is ${1\over 2}$
and for the last case the variance is 3.
\par
Figures 4--6 show the results of simulations
for both sets of segment lengths and all three processes.
For each scenario, the three
estimators were calculated at a range of distances
for $10{,}000$ simulations.
Generally speaking, $\hat K_R$ and $\Kvee_R$ behave similarly and
are superior to $\tilde K_R$, especially at longer distances when
the $Q_j$s are unequal.
Figure~4 shows the mean squared errors for $\hat K_R$.
In all cases, the contributions of the squared biases to the mean
squared errors are practically negligible and are always less than $0.5$\%.
As expected, the mean squared errors grow with $t$, especially
for the unequal segment length case as $t$ gets near~5, the longest
segment length available.
Another expected result is that the mean squared errors increase
with increasing clumpiness of the underlying process.
Figure~5 compares $\tilde K_R$ and $\hat K_R$.
We see that $\hat K_R$ is generally superior, although $\tilde K_R$ is sometimes
slightly better for smaller $t$.
The relative advantage of $\hat K_R$ (and $\Kvee_R$) over $\tilde K_R$
tends to be greater for more regular processes, which qualitatively agrees
with the asymptotic results in Stein (1995).
The advantage also tends to be greater for unequal segment lengths,
demonstrating that theoretical results obtained for equal segment lengths
may not accurately reflect the differences between estimators when
segment lengths are unequal.
Figure~6 compares $\hat K_R$ and $\Kvee_R$.  
{}From the theoretical results in the previous section, we should
expect these estimators to behave similarly when the waiting
time density is exponential so that the underlying model
is Poisson.
The simulations show that the estimators also tend to behave
very similarly for some non-Poisson models, especially when
the segment lengths are equal.
Neither estimator dominates the other, although $\Kvee$ tends to
be slightly superior for $t$ nearly as large as the longest segment length.
\par
For highly regular processes, $\hat K_R$ can be substantially inferior
to either $\tilde K_R$ or $\breve K_R$ for $t$ sufficiently small.
The problem is caused by the fact that in such circumstances,
having a pair of events within $t$ of each other is rare, so that
$\var\{T(\varphi)\}$ is much smaller than under a Poisson model
with the same intensity, whereas the variance of
$$
T(\varphi)-T^*(\varphi)
= {2(N_\bu-1)\over Q_\bu} \sum_{j=1}^{N_\bu}\{h(X_j,L_j;\varphi)
-Eh(X,L;\varphi)\}
$$ 
is not much different for a highly regular
process than for a Poisson process.
As a consequence, subtracting off $T(\varphi)-T^*(\varphi)$ from
$T(\varphi)$ tends to inflate the variance of the estimator.
As an example of a highly regular process, consider the
stationary renewal
process with waiting time density ${6^6\over 5!}x^5 e^{-x/6}$ for
$x>0$, a gamma density with parameters $6$ and ${1\over 6}$.
This waiting time distribution has mean~1 and variance ${1\over 6}$
and corresponds to a highly regular point process.
It is possible to show that
$$
K(t)=2t-{5\over 6}+{1\over 6}e^{-12t}+{1\over 3}\cos(3^{3/2}t)(e^{-9t}
+e^{-3t})+{1\over 3^{1/2}}\sin(3^{3/2}t)\left({1\over 3}e^{-9t}
+e^{-3t}\right)
$$
for this process.
Figure~7 shows that $\hat K_R$ is notably inferior to either $\tilde K_R$
and $\breve K_R$ for $t$ sufficiently small; for larger $t$, it is
competitive with $\breve K_R$ and clearly superior to $\tilde K_R$.
The overall winner is $\breve K_R$, which performs well for all $t$.
\par
We are unaware of any circumstances in which $\breve K_R$ performs
substantially worse than either $\hat K_R$ or $\tilde K_R$.
Thus, we recommend routinely using $\breve K_R$ to estimate
$K$, although routine adoption for processes in more than
one dimension will require the development of the necessary
software.
\par
\beginsection {7.} {Application to absorber catalog}
Figure~8 displays the estimators $\tilde K_R$, $\hat K_R$ and $\Kvee_R$
as applied to the absorber catalog described in Section~2.
The three estimators are very similar and, as expected, show
clear evidence of clustering of absorbers.
To obtain some idea about the uncertainty of these estimates, 
as in Quashnock and Stein (1999),
approximate 95\% pointwise confidence intervals were obtained
by bootstrapping using the 274 segments as the sampling units.
Specifically, using the notation in Section 5,
simulated absorber catalogs were produced by
sampling with replacement from $(Q_j;X_{1j},\ldots,X_{N_j j})$ for
$j=1,\ldots,274$, so that when one
selects a segment, one automatically selects the absorber locations
that go with this segment.
The confidence bands displayed in Figure~8 are then what Davison and Hinkley
(1997, p.~29) call the basic bootstrap confidence limits and are based
on 999 simulated catalogs.
All three estimators yield similar confidence intervals, which is disappointing
but perhaps not unexpected given the strong clustering that exists in 
the absorber catalog and the finding in the simulation study that the
advantage of the modifications decreases as clustering increases.
For these bootstrapping intervals to
be appropriate, $(Q_j;X_{1j},\ldots,X_{N_j j})$ for
$j=1,\ldots,274$ should be iid random objects.
Since the segments are of widely varying lengths, if the $Q_j$s are
viewed as fixed, the identically distributed assumption is false.
However, if we view the $Q_j$s as being a sequence of iid positive
random variables that are independent of the locations of absorbers,
then the identically distributed assumption may be reasonable.
Whether or not the independence assumption is reasonable depends
on the spatial extent of clustering among absorbers.
If there is no spatial dependence in absorber locations beyond, say,
100 $h^{-1}$ Mpc, then the independence assumption is not seriously
in error, since few pairs of segments are within this distance
of each other.
If, however, nonnegligible
clustering exists well beyond 100 $h^{-1}$ Mpc, then the independence
assumption is more problematic.

Analyses of galaxy surveys (Davis and Peebles 1983, Loveday, et al.\
1995) show that visible matter clusters on scales
of up to 20 $h^{-1}$ Mpc.
Thus, it is more interesting to investigate how
$K(t)-2t$ changes at distances beyond 20 $h^{-1}$ Mpc than to look
at $K$ itself.
Figure~8 shows that $\hat K_R(t)-2t$ generally increases until about
200 $h^{-1}$ Mpc and it is important to assess the uncertainty in
this pattern.
Applying the bootstrapping procedure to $\tilde K_R(t)-\tilde K_R(t_0)$
for $t_0=20, 50, 100$ and 150 $h^{-1}$ Mpc, Quashnock and Stein (1999)
concluded that there was strong evidence for clustering from
20 to 50 $h^{-1}$ Mpc and from 50 to 100 $h^{-1}$ Mpc, but at best
marginal evidence for clustering beyond 100 $h^{-1}$ Mpc.
The results with the modified
estimates (not shown) confirm the clear evidence for clustering from
20 to 50 $h^{-1}$ Mpc and from 50 to 100 $h^{-1}$ Mpc.
Figure~9 shows the lower bounds for pointwise 95\% confidence intervals
for $K(t)-K(100)-2(t-100)$.
The modified estimators yield slightly stronger evidence of clustering
beyond 100 $h^{-1}$ Mpc, which is mostly due to the fact that the
modified estimates of $K(t)-K(100)-2(t-100)$ are slightly larger
than the unmodified estimates for $t$ around 200 and not because
the modified intervals are narrower.
If one used 99\% pointwise confidence intervals in Figure~9, then
for all $t>100$ and all three estimators, the lower confidence
bounds are negative.
Thus, the conclusion in Quashnock and Stein (1999) that there is 
perhaps marginal evidence for clustering beyond 100 $h^{-1}$ Mpc is
not altered by using the modified estimators.
\par
As discussed in Section 2, the broad range of redshifts in the absorber
catalog implies that we are looking at the universe at a broad range
of times.
The use of comoving units largely equalizes the intensity of absorbers
across redshifts, but it does not equalize the clustering.
Indeed, by dividing the absorber catalog into groups based on their
redshift, Quashnock and Vanden Berk (1998) found evidence that as redshift
decreases, clustering on the scales of 1 to 16 $h^{-1}$ Mpc
strongly increases across the range of redshifts in the absorber
catalog.
Quashnock and Vanden Berk (1998) further note that this
increase in clustering with decreasing redshift is
consistent with what is known through theory and simulations about
how gravity should affect the evolution of the clustering of
absorbers over time.
Using the various forms of the rigid motion estimator of $K$ described
here on groups of the absorber catalog with similar redshifts, we also find
that on the scale of a few tens of $h^{-1}$ Mpc,
clustering increases substantially with decreasing redshift
over the range of redshifts in the absorber catalog (results not
shown).
Thus, on these shorter scales,
our estimates of $K$ measure an average clustering
over the range of redshifts in the absorber catalog.
\par
In contrast,
Quashnock, Vanden Berk and York (1996) found no evidence that
clustering at scales of 100 $h^{-1}$ Mpc changes over the redshift
range in the absorber catalog.
Similarly, when looking at, say, $\breve K_R(t)-\breve K_R(100)$
for $t>100$ based on higher and lower redshift parts of the catalog,
we find no systematic difference in the estimates as a function
of redshift.
For example, dividing the 274 segments in the catalog into two groups
of size 137 based on redshift, $\breve K_R(150)-\breve K_R(100)$
equals 150.8 for the lower redshift group and 151.4 for the higher
redshift group.
Thus, we do not believe that the modest evidence we find for clustering
at these larger scales is due to
inhomogeneities across time in the distribution of absorbers.
\par
\beginsection {8.} {Summary}
For studying the behavior of edge-corrected estimators of the $K$ function
of a point process, taking the observation domain to be a sequence
of segments has a number of desirable consequences.
First, explicit expressions are available for a number
of the more popular estimators, which is often not the case for regions
in more than one dimension.
The availability of such explicit expressions eases the study of the
properties of these estimators via both theory and simulation. 
In addition, studying settings in which the number of segments is
large yields results that highlight the differences between the 
various methods of edge-correction.
In particular, simulation results show that
allowing the segment lengths to vary generally  
increases the differences between estimators.
The overall conclusion about the merits of the various estimators
is that $\breve K_R$, a modification of the rigid motion estimator
based on an approach suggested by Picka (1996), is 
the estimator of choice.
\par
The absorber catalog studied here shows that multiple windows of
varying size can arise in practice.
Although it is somewhat disappointing that the bootstrap confidence
intervals for the ordinary rigid motion corrected estimator and its
modifications are very similar, this result is not too surprising
in light of the simulation results showing that the benefit of
the modifications is smaller for clustered processes.
The simulation results indicate that the modified estimators
can have substantially smaller mean squared errors for Poisson
or more regular processes, especially if the segment lengths
vary substantially.
\par
\beginsection {Appendix.} {Proofs}
We first derive (\Eqn 2/) assuming, for convenience,
the $Q_j$s have been arranged
in increasing order.  We have
$$\eqalignno{
{1\over Q_\bu}h(x,\ell;\varphi^R) 
 &= {1\over Q_\sbullet}\sum_{j=1}^p\int_0^{Q_j}\varphi^R
     \big((x,\ell),(y,j)\big)dy\cr
 &= \int_0^{Q_\ell} {1\{|x-y|\le t\} \over U(|x-y|)}\,
     dy\cr
 &= \int_0^x {1\{|x-y|\le t\} \over U(|x-y|)}\,
     dy + \int_x^{Q_\ell} {1\{|x-y|\le t\} \over 
     U(|x-y|)}\, dy\cr
 &= \int_0^x {1\{|x-y|\le t\} \over U(|x-y|)}\,
     dy + \int_0^{Q_\ell-x} {1\{|Q_\ell-x-y|\le t\} \over 
     U(|Q_\ell-x-y|)}\, dy.\cr}
$$
Thus, to verify (\Eqn 2/), we need to show that
$$
\kappa(x,t)= \int_0^x {1\{|x-y|\le t\} \over U(|x-y|)}\, dy.
$$
Now
$$
\eqalignno{
\int_0^x {1\{|x-y|\le t\} \over U(|x-y|)}\, dy 
 &= \int_{(x-t)^+}^x {dy \over U(x-y)}  \cr
 &= \sum_{k=1}^{j(x\land t)-1}\int_{x-Q_k}^{x-Q_{k-1}}
  {dy \over \sum_{j=k}^p (Q_j-x+y)} \cr
  &\qquad + \int_{(x-t)^+}^{x-Q_{j(x\land t)-1}} 
   {dy \over \sum_{j=j(x\land t)}^p
   (Q_j-x+y)},\cr}
$$
which equals $\kappa(x,t)$ by calculus.
\par
We next derive $S_2(\varphi^R)$, again assuming the $Q_j$s have been arranged
in increasing order.
By the symmetry of $\varphi^R$,
$$
S_2(\varphi^R)=2Q_\sbullet^2\sum_{j=1}^p
\int_0^{Q_j}\int_0^x {1\{x-y\le t\}\over
U(x-y)^2}\, dy\,dx,
$$
so taking $v=x-y$ and then switching the order of integration yields
$$\eqalignno{
{S_2(\varphi^R)\over 2Q_\sbullet^2} 
&= \sum_{j=1}^p \int_0^{Q_j} \int_0^{x\land t}
{1\over U(v)^2}\, dv\, dx\cr
&=  \sum_{j=1}^p \int_0^{Q_j\land t} {Q_j-v\over U(v)^2}
    \,dv\cr
&=  \sum_{j=1}^p \sum_{\ell=1}^{j\land \{j(t)-1\}}
    \int_{Q_{\ell-1}}^{Q_\ell} {Q_j-v\over \left\{\sum_{k=\ell}^p
    (Q_k-v)\right\}^2}\, dv
 + \sum_{j=j(t)}^p \int_{Q_{j(t)-1}}^t {Q_j-v\over
     \big\{\sum_{k=j(t)}^p (Q_k-v)\big\}^2}\, dv\cr
&= \sum_{j=1}^p \sum_{\ell=1}^{j\land \{j(t)-1\}}
   \left\{ {Q_j-Q_\ell\over (p-\ell+1)U_\ell} - {Q_j-Q_{\ell-1}\over
   (p-\ell+1)U_{\ell-1}} - {1\over (p-\ell+1)^2}\log\left(
   {U_\ell\over U_{\ell-1}}\right)\right\}\cr
&\qquad + \sum_{j=j(t)}^p
   \left[ {Q_j-t\over \{p-j(t)+1\}U(t)} - {Q_j-Q_{j(t)-1}\over
   \{p-j(t)+1\}U_{j(t)-1}} - {1\over \{p-j(t)+1\}^2}\log\left\{
   {U(t)\over U_{j(t)-1}}\right\}\right].\cr
}
$$
Using the definition of $U(t)$, the second sum simplifies to
$\{p-j(t)+1\}^{-1}\log\{U_{j(t)-1}/U(t)\}$ and by switching
the order of summation and using the definition of $U_\ell$,
the first sum equals
$$
\eqalignno{
& \sum_{\ell=1}^{j(t)-1}\sum_{j=\ell}^p
   \left\{ {Q_j-Q_\ell\over (p-\ell+1)U_\ell} - {Q_j-Q_{\ell-1}\over
   (p-\ell+1)U_{\ell-1}} + {1\over (p-\ell+1)^2}\log\left(
   {U_{\ell-1}\over U_{\ell}}\right)\right\}\cr
& \quad = 
\sum_{\ell=1}^{j(t)-1} \left\{ 
 {U_\ell\over (p-\ell+1)U_\ell} 
  - {U_{\ell-1}\over(p-\ell+1)U_{\ell-1}} +{1\over p-\ell+1}
  \log\left({U_{\ell-1}\over U_{\ell}}\right)\right\}\cr
& \quad = \sum_{\ell=1}^{j(t)-1} {1\over
   p-\ell+1}\log\left({U_{\ell-1}\over U_{\ell}}\right).\cr
}
$$
Thus,
$$
S_2(\varphi^R)= 2Q_\sbullet^2 \sum_{\ell=1}^{j(t)-1} {1\over
   p-\ell+1}\log\left({U_{\ell-1}\over U_{\ell}}\right)
   + {2Q_\sbullet^2 \over p-j(t)+1}
   \log\left\{{U_{j(t)-1}\over U(t)}\right\}.
$$
If $Q_1=\cdots=Q_p=Q$, then for $t<Q$, $j(t)=1$, so
$$
S_2(\varphi^R)=-2pQ^2\log\left(1-{t\over Q}\right).\eqno(\SNM 6/)
$$
\par
Calculating $S_1(\varphi^R)$ is more difficult and we only
give the special case $Q_1=\cdots=Q_p=Q$.
Setting $s=t/Q$, we then have
$$
S_1(\varphi^R) = p \int_0^Q\left\{\int_0^Q
 {Q1\{|x-y|\le t\}\over Q-|x-y|}dy\right\}^2\! dx
= 4pQ^3\gamma(s),\eqno(\SNM S1.form/)
$$
where $\gamma$ is defined in (\Eqn gamma.def/).
To evaluate $\gamma$, write
$$\eqalignno{
\gamma(s) & = {1\over 2}\int_0^1\left[\int_0^x {1\{x-y\le s\}\over 1-x+y}
  \,dy\right]^2 dx\cr
&\qquad +{1\over 2}\int_0^1\left[\int_0^x {1\{x-y\le s\}\over 1-x+y}\,dy\right]
 \left[\int_x^1 {1\{z-x\le s\}\over 1-z+x}\,dz\right]dx  \cr
&= {1\over 2}\int_0^1\log^2\{ 1-(x\land s)\}\,dx
  +{1\over 2}\int_0^1\log\{ 1-(x\land s)\}
  \log\{ (1-s)\lor x\}\,dx.\cr}
$$
Now
$$
\int_0^1\log^2\{ 1-(x\land s)\}dx=
  2s+2(1-s) \log (1-s) 
$$
and for $s\le{1\over 2}$,
$$
 \int_0^1\log\{ 1-(x\land s)\}
  \log\{ (1-s)\lor x\}dx
= -\log^2(1-s)-2s\log(1-s)
$$
whereas for $s>{1\over 2}$,
$$
\eqalignno{
 &\int_0^1\log\{ 1-(x\land s)\}
  \log\{ (1-s)\lor x\}\,dx\cr
 &\quad = -2(1-s)\log(1-s)
  -2s \log s \log (1-s) + \int_{1-s}^s \log(1-y)\log y\, dy.\cr}
$$
Hence,
$$\eqalignno{
\gamma(s)=
s & +(1-2s)^+\log(1-s) -1\left\{s\le {1\over 2}\right\}{1\over 2}
\log^2(1-s)\cr
 & -1\left\{s>{1\over 2}\right\}s\log s\log(1-s)
   +\int_0^{(s-1/2)^+}\log\left({1\over 2} -y
\right)\log\left({1\over 2}+y\right)dy. &(\SNM gamma/) \cr}
$$
\par
Let us next consider computing $S_2(\varphi^I)$.
Defining $R(v)=Q_\sbullet-\sum_{j=1}^p(2v-Q_j)^+$, then for $y<x<Q_\ell$
we have
$$
\varphi^I\big((x,\ell),(y,m)\big)= {1\{x-y\le t,\ell=m\}Q_\sbullet\over
  R(x-y)}\left[{1\over 1+ 1\{2x-y<Q_\ell\}} + {1\over 1+ 1\{2y-x>0\}}
  \right].
$$
Thus, taking $v=x-y$,
$$
\eqalignno{
{S_2(\varphi^I)\over 2Q_\sbullet^2} &= \sum_{\ell=1}^p \int_0^{Q_\ell}
  \!\int_0^x {1\{x-y\le t\}\over R(x-y)^2}
  \left[{1\over 1+ 1\{2x-y<Q_\ell\}} + {1\over 1+ 1\{2y-x>0\}}
  \right]^2 dy\, dx\cr
&= \sum_{\ell=1}^p \int_0^{Q_\ell}\! \int_0^{x\land t} {1\over R(v)^2}
    \left[{1\over 1+ 1\{x+v<Q_\ell\}} + {1\over 1+ 1\{x>2v\}}
  \right]^2 dv\, dx\cr
&= \sum_{\ell=1}^p \int_0^{t\land Q_\ell}{1\over R(v)^2}\int_v^{Q_\ell}
    \left[{1\over 1+ 1\{x+v<Q_\ell\}} + {1\over 1+ 1\{x>2v\}}
  \right]^2 dx\, dv.\cr}
$$
Now $[1+ 1\{x+v<Q_\ell\}]^{-1}+ [ 1+ 1\{x>2v\}]^{-1}$
takes on values $2$, ${3\over 2}$ and 1 depending on, 
respectively, whether none, one or both of
$x+v<Q_\ell$ and $x>2v$ are true.  
Thus,
$$
\eqalignno{
{S_2(\varphi^I)\over 2Q_\sbullet^2} &=
 \sum_{\ell=1}^p\left\{\int_0^{t\land {1\over 3}Q_\ell}
  {{9\over 4}2v+ 1(Q_\ell-3v)\over R(v)^2}\,dv\right.\cr
&\quad\qquad\left. +\int_{t\land {1\over 3}Q_\ell}^{t\land{1\over 2}Q_\ell}
  {{9\over 4}(2Q_\ell-4v)+ 4(3v-Q_\ell)\over R(v)^2}\,dv
 +\int_{t\land {1\over 2}Q_\ell}^{t\land Q_\ell}
  {4(Q_\ell-v)\over R(v)^2}\,dv\right\}\cr
&= \sum_{\ell=1}^p\left\{\int_0^{t\land {1\over 3}Q_\ell} {Q_\ell+
  {3\over 2}v\over
  R(v)^2}\,dv + \int_{t\land {1\over 3}Q_\ell}^{t\land{1\over 2}Q_\ell}
  {{1\over 2}Q_\ell+3v\over R(v)^2}\,dv + \int_{t\land {1\over
2}Q_\ell}^{t\land Q_\ell}{4(Q_\ell-v)\over R(v)^2}\,dv\right\}.\cr}
$$
\par
While it is possible to evaluate these integrals explicitly, the 
resulting expressions do not appear to simplify as in the case for
the rigid motion estimator.
When $Q_1=\cdots=Q_p=Q$, we do obtain a fairly simple
explicit result.
By taking $u=v/Q$, we get
$$
\eqalignno{
S_2(\varphi^I) &=2pQ^2\left[ \int_0^{s\land{1\over 3}}
 {1+{3\over 2}u\over\{1-(2u-1)^+\}^2}\, du +\int_{s\land {1\over 3}}^{s
  \land{1\over 2}}{{1\over 2}+3u\over\{1-(2u-1)^+\}^2}\, du\right.\cr
&\left.\qquad\qquad + \int_{s\land{1\over 2}}^s {4-4u\over
 \{1-(2u-1)^+\}^2}\, du\right]\cr
&=2pQ^2\left\{ \int_0^{s\land {1\over 3}}
(1+{3\over 2}u)\, du+ \int_{s\land {1\over 3}}^{s\land{1\over
 2}}\left({1\over 2}+3u\right) du
 + \int_{s\land{1\over 2}}^s {1\over 1-u}\, du
\right\},\cr
}$$
so that for $s=t/Q<1$,
$$
S_2(\varphi^I)= 2pQ^2\times \cases
{  s +{3\over 4}s^2 & if $0<s\le {1\over 3}$,\cr
 {1\over 12}+ {1\over 2}s+ {3\over
  2}s^2& if ${1\over 3}\le s\le {1\over 2}$ and\cr
{17\over 24} - \log 2 -\log(1-s)& 
   if ${1\over 2}\le s<1$.\cr}\eqno(\SNM 8/)
$$
\par
\beginsection {References}
\beginref
Baddeley, A. (1998). Spatial sampling and censoring.
In {\it Stochastic Geometry: Likelihood and Computation\/}
(O.\ E.\ Barndorff-Nielsen, W.\ S.\ Kendall and
M.\ N.\ M.\ van Lieshout, eds.) Chapter 2.
Chapman and Hall, London.

Baddeley, A.\ and Gill, R.\ D. (1997). Kaplan-Meier estimators of
distance
distributions for spatial point processes. {\it Ann.\ Statist.\/}
{\bf 25} 263--292.

Baddeley, A.\ J., Moyeed, R.\ A., Howard, C.\ V.\ and Boyde, A. (1993).
Analysis of a three-dimensional point pattern with replication.
{\it Appl.\ Statist.\/} {\bf 42} 641--668.

Coleman, P.\ H.\ and Pietronero, L. (1992). The fractal structure
of the universe. {\it Phys.\ Reports\/} {\bf 213} 311--389.

Crotts, A.\ P.\ S., Melott, A.\ L.\ and York, D.\ G. (1985).
QSO metal-line absorbers:  the key to large-scale structure?
{\it Phys.\ Letters B\/} {\bf 155B} 251--254.

Daley, D.\ J.\ and Vere-Jones, D. (1988). {\it An Introduction to the
Theory of Point Processes.\/} Springer-Verlag, New York.

Davis, M.\ and Peebles, P.~J.~E. (1983).  A survey of galaxy redshifts.
V. The two-point position and velocity correlations.
{\it Astrophys.\ J.\/} {\bf 267} 465--482.

Davison, A.\ C.\ and Hinkley, D.\ V. (1997). {\it Bootstrap Methods
and Their Application.\/} Cambridge University Press.

Feller, W. (1971). {\it An Introduction to Probability Theory and
Its Applications\/}, vol.~II. Wiley, New York.

Hamilton, A.\ J.\ S. (1993). Toward better ways to measure the galaxy
correlation function. {\it Astrophys.\ J.\/} {\bf 417} 19--35.

Jing, Y.~P.\ and Suto, Y. (1998).  Confronting cold dark matter cosmologies
with strong clustering of Lyman break galaxies at $z\sim 3$.
{\it Astrophys.\ J.\/} {\bf 494} L5--L8.

Landy, S.\ L.\ and Szalay, A.\ S. (1993). Bias and variance of 
angular correlation functions. {\it Astrophys.\ J.\/} {\bf 412} 64--71.

Lanzetta, K.~M., Bowen, D.~B., Tytler. D.\ and Webb, J.~K. (1995).
The gaseous extent of galaxies and the origin of Lyman-alpha absorption
systems: A survey of galaxies in the fields of
Hubble Space Telescope spectroscopic target QSOs.
{\it Astrophys.\ J.\/} {\bf 442} 538--568.

Loveday, J., Maddox, S.~J., Efstathiou, G.\ and Peterson, B.~A. (1995).
The Stromlo-APM redshift survey. II. Variation of galaxy clustering with
morphology and luminosity. {\it Astrophys.\ J.\/} {\bf 442} 457--468.

Margon, B. (1999).  The Sloan Digital Sky Survey.
{\it Phil.\ Trans.\ R.\ Soc.\ Lond.\ A\/} {\bf 357} 93-103.

Mart\'{\i}nez, V.\ J. (1997). Recent advances in large-scale structure
statistics. In {\it Statistical Challenges in Modern Astronomy II\/}
(G.\ J.\ Babu and E.\ D.\ Feigelson, eds.) 153--166. Springer, New
York.

Ohser, J. (1983). On estimators for the reduced second moment measure of
point processes. {\it Math.\ Oper.\ Statist.\ ser Statist.\/} {\bf 14}
63--71.

Ohser, J.\ and Stoyan, D. (1981).  On the second-order and
orientation analysis of planar stationary point processes.
{\it Biom.\ J.\/} {\bf 23} 523--533.

Peebles, P.\ J.\ E. (1993). {\it Principles of Physical Cosmology\/}.
Princeton University Press.

Picka, J. (1996).  Variance-reducing modifications for estimators of
dependence in random sets.  Ph.\ D.\ dissertation, Department of Statistics,
University of Chicago.

Pons-Border\'{\i}a, M.-J., Mart\'{\i}nez, V.\ J., Stoyan, D., Stoyan, H.
and Saar, E. (1999).  Comparing estimators of the galaxy correlation
function.  {\it Astrophys.\ J.\/} {\bf 523} 480--491.

Quashnock, J.\ M.\ and Stein, M.\ L. (1999).
A measure of clustering of QSO heavy-element absorption-line systems.  
{\it Astrophys.\ J.\/} {\bf 515} 506--511.

Quashnock, J.\ M.\ and Vanden Berk, D.\ E. (1998). The form and
evolution of the clustering of QSO heavy-element absorption-line
systems. {\it Astrophys.\ J.\/} {\bf 500} 28--36.

Quashnock, J.\ M., Vanden Berk, D.\ E.\ and York, D.\ G. (1996).
High-redshift superclustering of quasi-stellar object absorption-line
systems on 100 $h^{-1}$ Mpc scales.
{\it Astrophys.\ J.\/} {\bf 472} L69-L72.

Ripley, B.\ D. (1988). {\it Statistical Inference for Spatial
Processes.\/} Cambridge University Press, Cambridge.

Steidel, C.~C., Adelberger, K.~L., Dickinson, M., Giavalisco, M. Pettini,
M.\ and Kellogg, M. (1998).
A large structure of galaxies at redshift $z\sim 3$ and its cosmological
implications.
{\it Astrophys.\ J.\/} {\bf 492} 428--438.

Stein, M.\ L. (1993). Asymptotically optimal estimation for the reduced
second moment measure of point processes. {\it Biometrika\/} {\bf 80}
443--449.

Stein, M.\ L. (1995). An approach to asymptotic inference for spatial
point processes. {\it Statist.\ Sinica\/} {\bf 5} 221--234.

Stoyan, D.\ and Stoyan, H. (2000). Improving ratio estimators
of second order point process characteristics.  {\it Scand.\ J.\ Statist.\/} in
press.

Sylos Labini, F., Montuori, M.\ and Pietronero, L. (1998).
Scale-invariance of galaxy clustering.
{\it Phys.\ Reports\/} {\bf 293} 61--226.

Vanden Berk, D.\ E., Quashnock, J.\ M., York, D.\ G., Yanny, B. (1996).
An excess of  C~\IV absorbers in luminous quasars: evidence for
gravitational lensing? {\it Astrophys.\ J.\/} {\bf 469} 78--83.

York, D.~G., Yanny, B., Crotts, A., Carilli, C., Garrison, E.\ and
Matheson, L. (1991).
An inhomogeneous reference catalogue of identified intervening heavy element
systems in spectra of QSOs.
{\it Mon.\ Not.\ Roy.\ Astron.\ Soc.\/} {\bf 250} 24--49.

Zhang, Y., Meiksin, A., Anninos, P.\ and Norman, M.~L. (1998).
Physical properties of the Ly$\alpha$ forest in cold dark matter
cosmology. {\it Astrophys.\ J.\/} {\bf 495} 63--79.

\endref
\par

\vfill\eject
\epsfbox{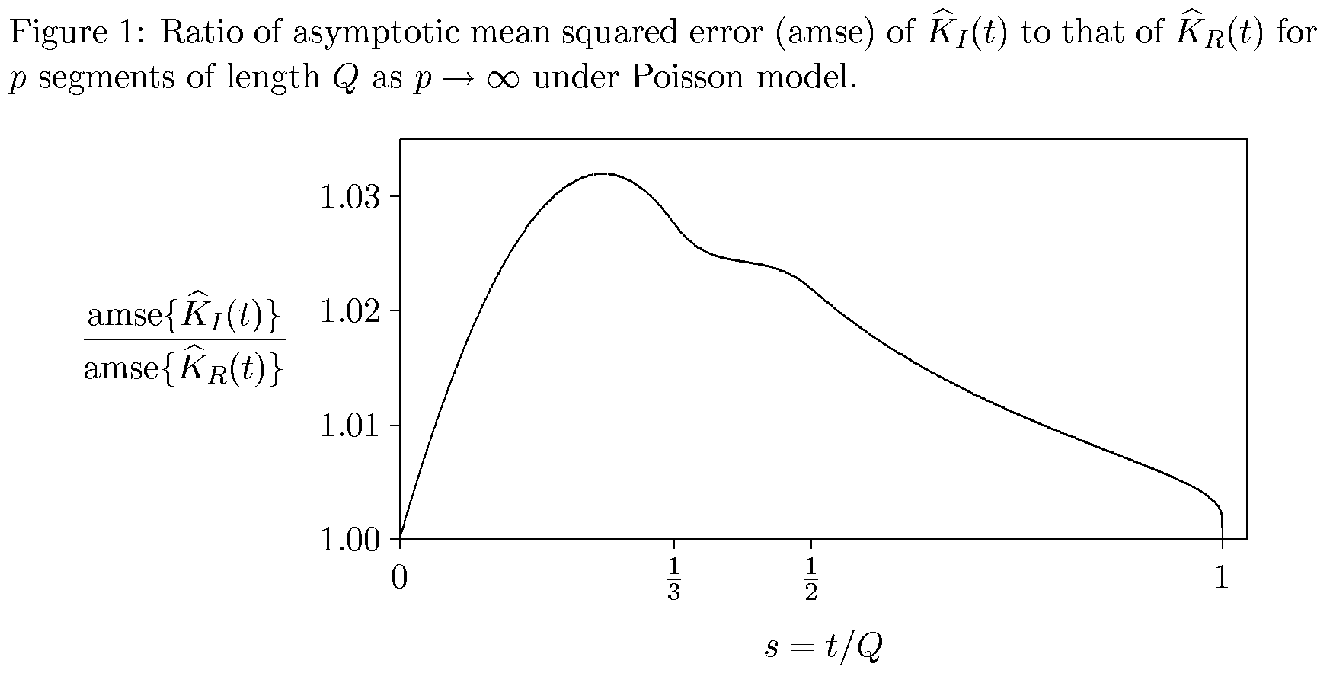}

\vfill\eject
\epsfbox{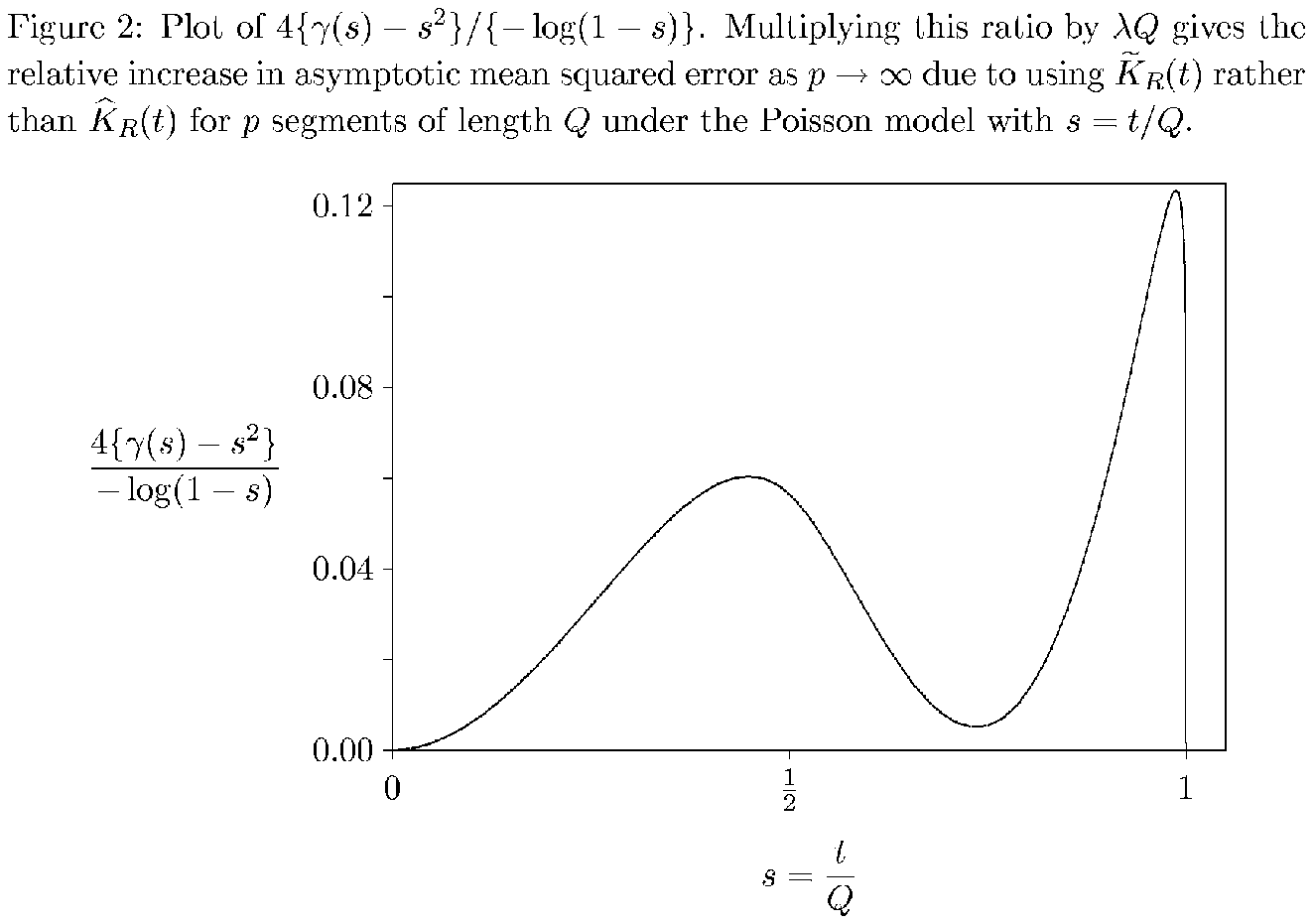}

\vfill\eject
\epsfbox{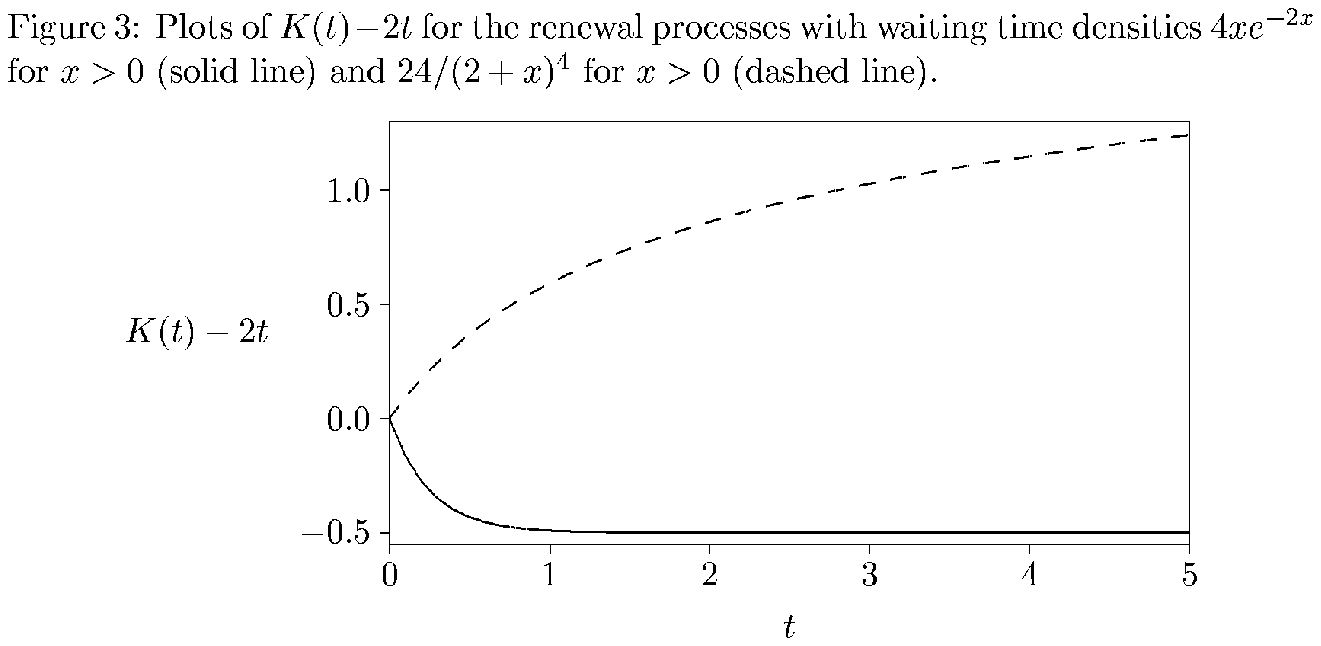}

\vfill\eject
\epsfbox{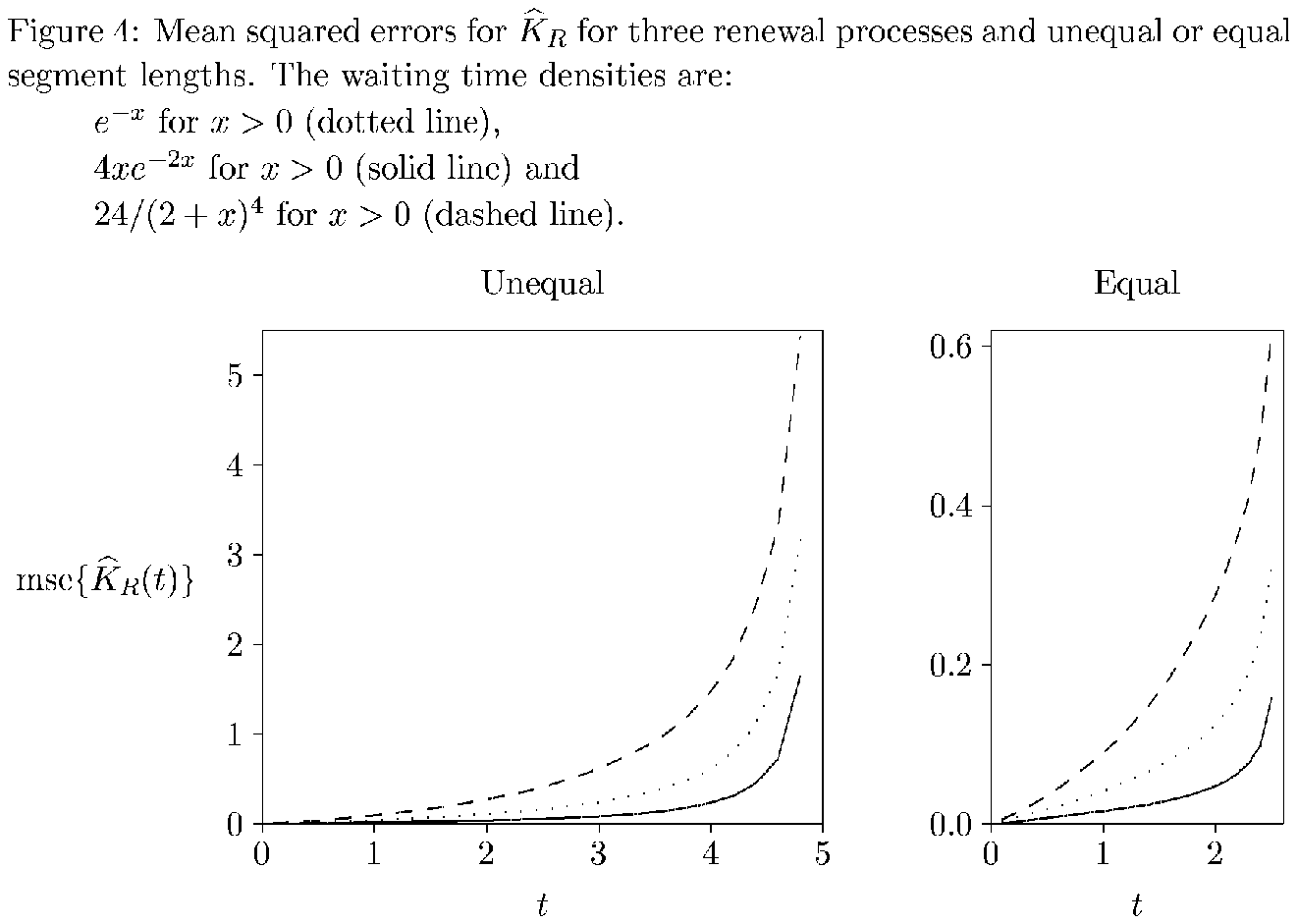}

\vfill\eject
\epsfbox{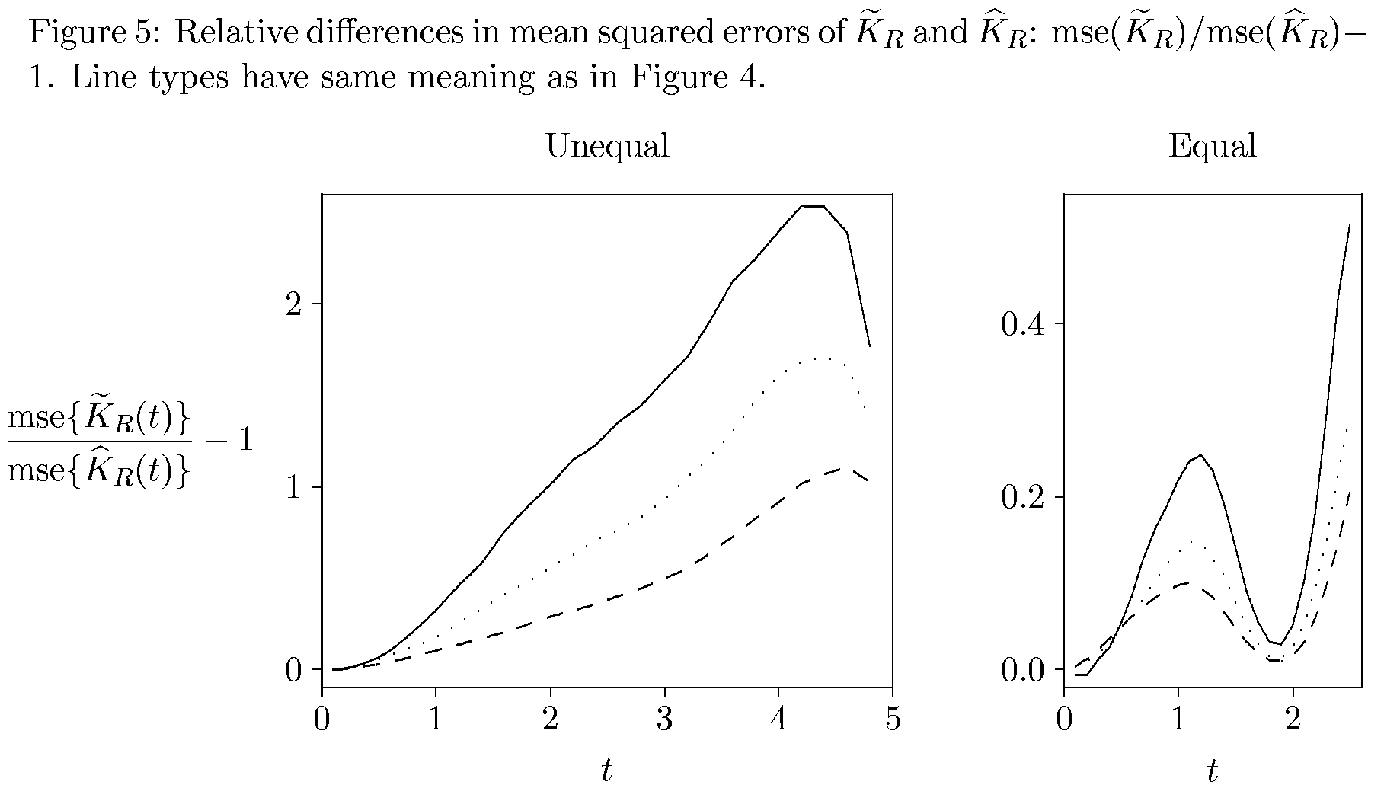}

\vfill\eject
\epsfbox{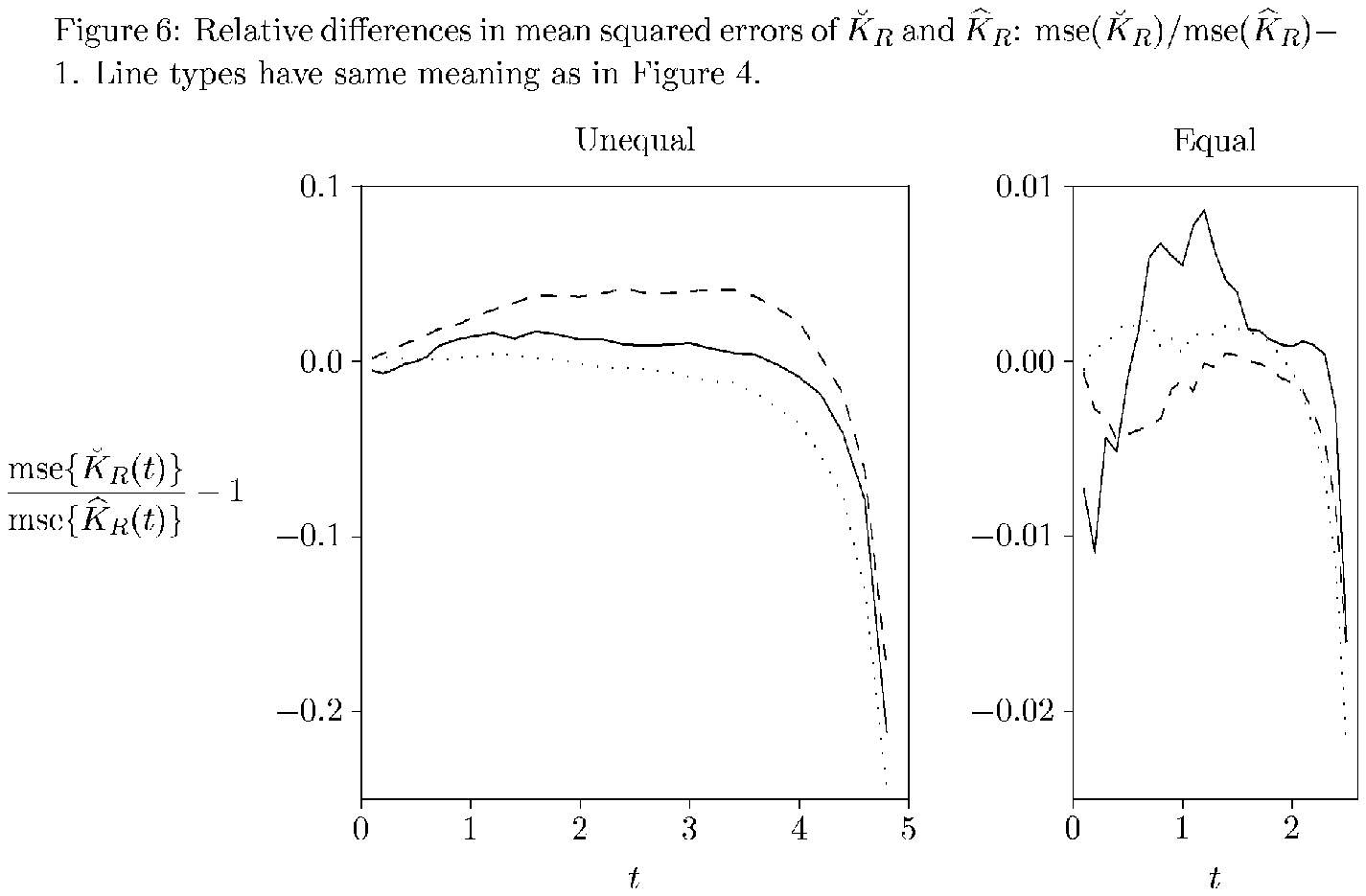}

\vfill\eject
\epsfbox{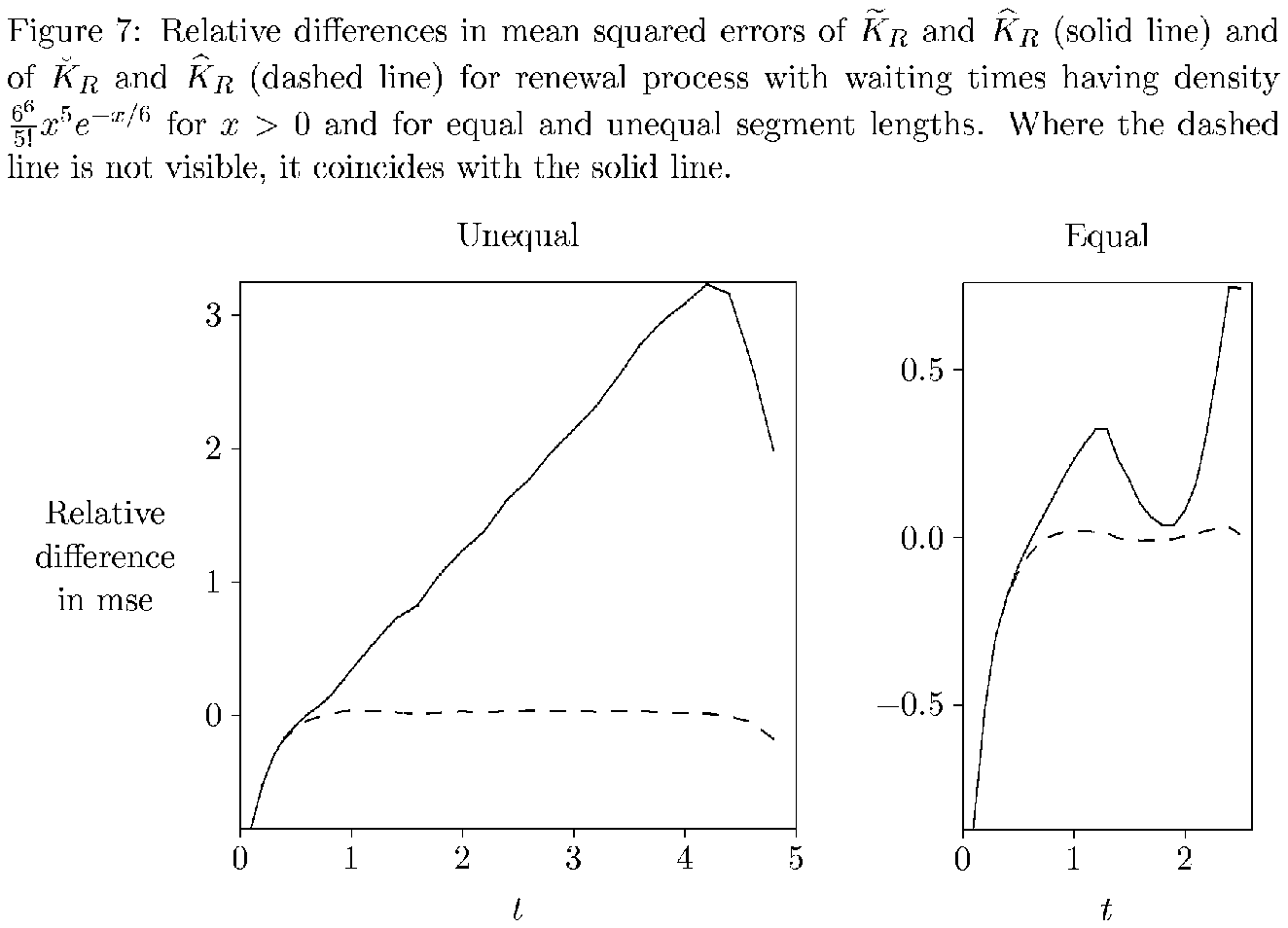}

\vfill\eject
\epsfbox{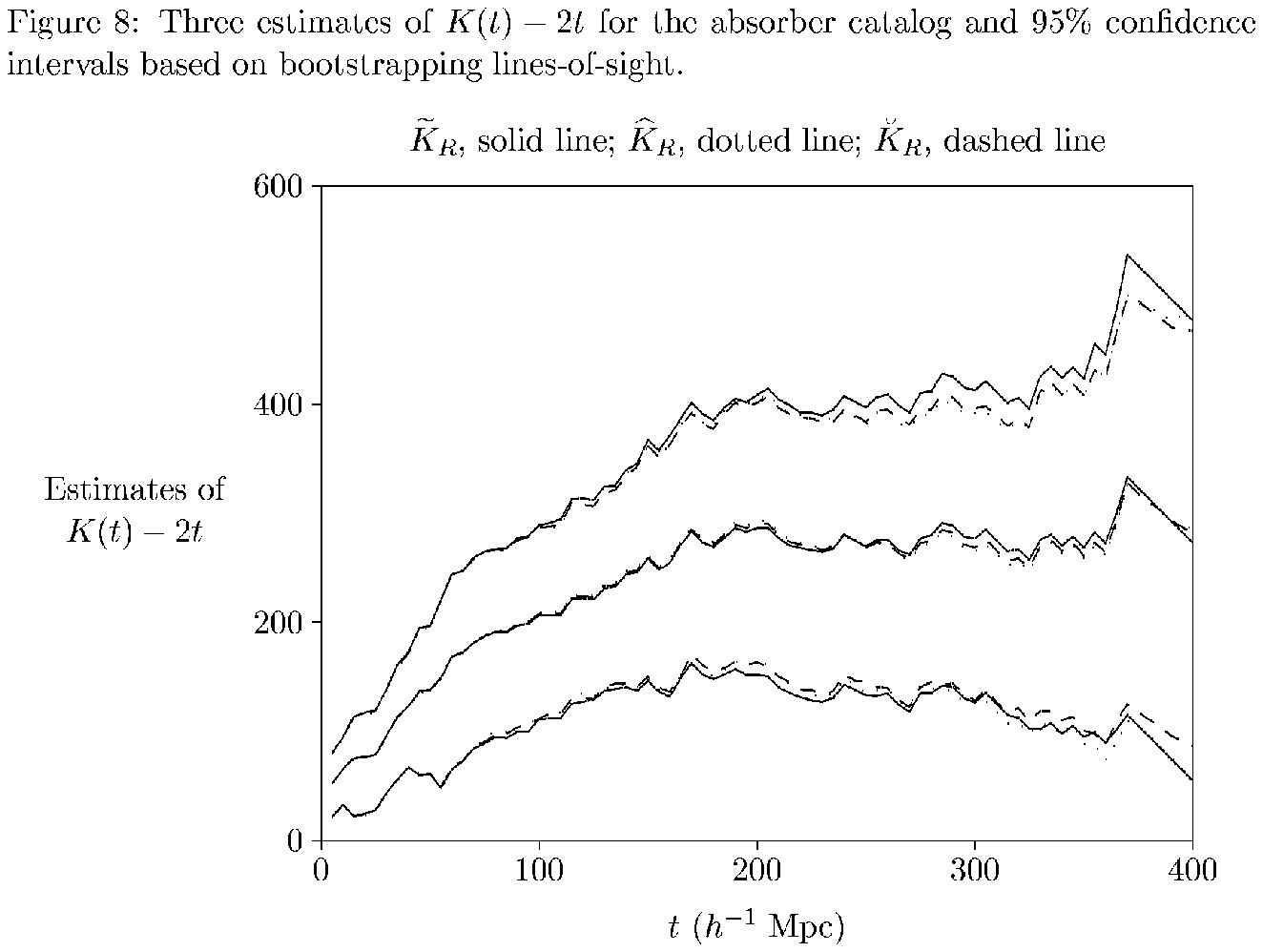}

\vfill\eject
\epsfbox{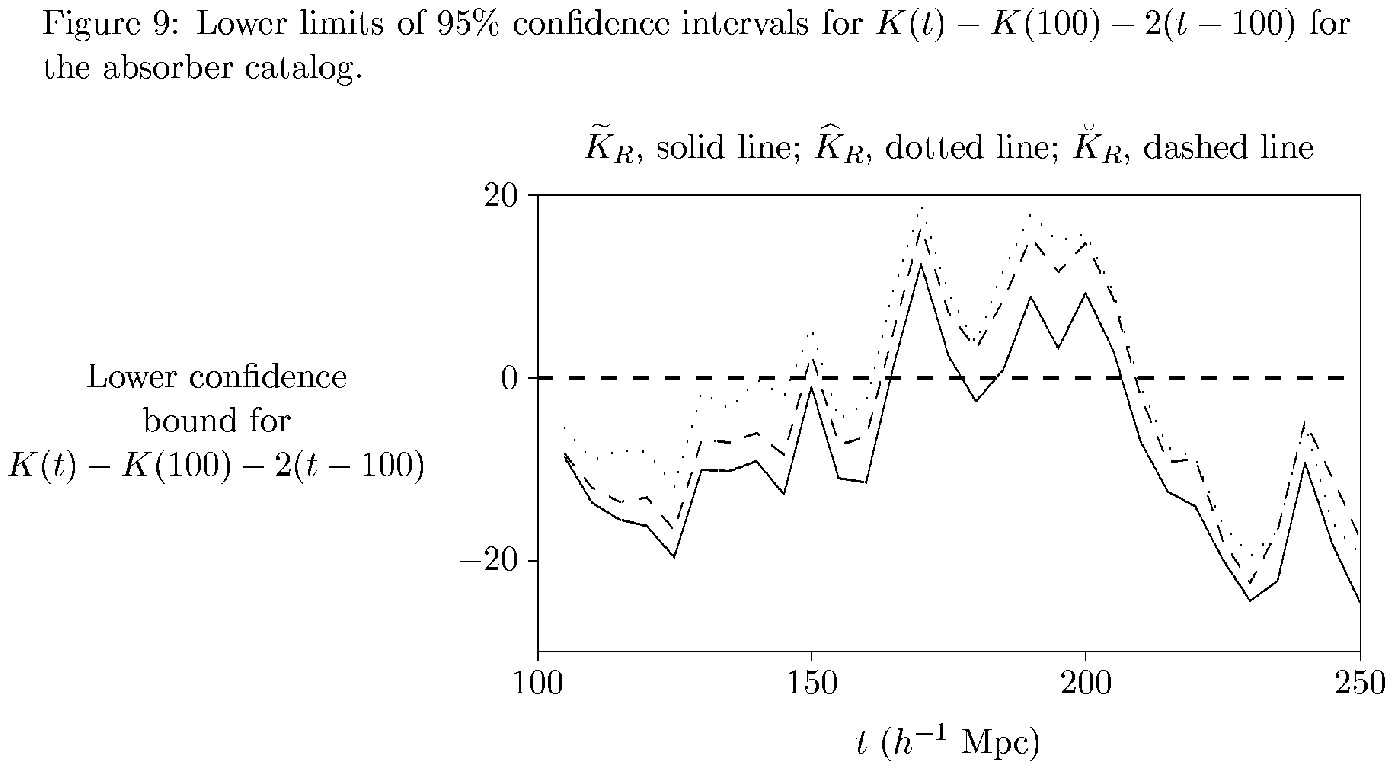}

\bye